\documentclass[12pt]{article}

\usepackage{amsmath,amssymb,amsfonts,cite, epsfig,setspace,bigstrut,framed}
\usepackage[all]{xy}
\usepackage{color}
\usepackage{pifont}
\usepackage{slashed}

\makeatletter \@addtoreset{equation}{section} \makeatother
\renewcommand{\theequation}{\thesection.\arabic{equation}}

\begin{document}

\vskip 0.25in

\newcommand{\todo}[1]{{\bf\color{blue} !! #1 !!}\marginpar{\color{blue}$\Longleftarrow$}}
\newcommand{\nn}{\nonumber}
\newcommand{\comment}[1]{}
\newcommand\T{\rule{0pt}{2.6ex}}
\newcommand\B{\rule[-1.2ex]{0pt}{0pt}}

\newcommand{\cM}{{\cal M}}
\newcommand{\cW}{{\cal W}}
\newcommand{\cN}{{\cal N}}
\newcommand{\cH}{{\cal H}}
\newcommand{\cK}{{\cal K}}
\newcommand{\cT}{{\cal T}}
\newcommand{\cZ}{{\cal Z}}
\newcommand{\cO}{{\cal O}}
\newcommand{\cQ}{{\cal Q}}
\newcommand{\cB}{{\cal B}}
\newcommand{\cC}{{\cal C}}
\newcommand{\cD}{{\cal D}}
\newcommand{\cE}{{\cal E}}
\newcommand{\cF}{{\cal F}}
\newcommand{\cX}{{\cal X}}
\newcommand{\cL}{{\cal L}}
\newcommand{\cP}{{\cal P}}
\newcommand{\cI}{{\cal I}}
\newcommand{\IA}{\mathbb{A}}
\newcommand{\IP}{\mathbb{P}}
\newcommand{\IQ}{\mathbb{Q}}
\newcommand{\IH}{\mathbb{H}}
\newcommand{\IR}{\mathbb{R}}
\newcommand{\IC}{\mathbb{C}}
\newcommand{\IF}{\mathbb{F}}
\newcommand{\IV}{\mathbb{V}}
\newcommand{\II}{\mathbb{I}}
\newcommand{\IZ}{\mathbb{Z}}
\newcommand{\re}{{\rm Re}}
\newcommand{\im}{{\rm Im}}
\newcommand{\tr}{\mathop{\rm Tr}}
\newcommand{\ch}{{\rm ch}}
\newcommand{\rk}{{\rm rk}}
\newcommand{\ext}{{\rm Ext}}
\newcommand{\bi}{\begin{itemize}}
\newcommand{\ei}{\end{itemize}}
\newcommand{\beq}{\begin{equation}}
\newcommand{\eeq}{\end{equation}}
\newcommand{\bea}{\begin{eqnarray}}
\newcommand{\eea}{\end{eqnarray}}
\newcommand{\ba}{\begin{array}}
\newcommand{\ea}{\end{array}}

\newcommand{\CA}{\mathbb A}
\newcommand{\CP}{\mathbb P}
\newcommand{\tmat}[1]{{\tiny \left(\begin{matrix} #1 \end{matrix}\right)}}
\newcommand{\mat}[1]{\left(\begin{matrix} #1 \end{matrix}\right)}
\newcommand{\diff}[2]{\frac{\partial #1}{\partial #2}}
\newcommand{\gen}[1]{\langle #1 \rangle}

\newtheorem{theorem}{\bf THEOREM}
\newtheorem{proposition}{\bf PROPOSITION}
\newtheorem{observation}{\bf OBSERVATION}

\def\theequation{\thesection.\arabic{equation}}
\newcommand{\setall}{
	\setcounter{equation}{0}
}
\renewcommand{\thefootnote}{\fnsymbol{footnote}}

\begin{titlepage}
\vfill
\begin{flushright}
{\tt\normalsize KIAS-P13056}\\

\end{flushright}
\vfill
\begin{center}
{\Large\bf Abelianization of BPS Quivers \\ \vskip 3mm and the Refined Higgs Index }

\vskip 1cm
Seung-Joo Lee${}^1$\footnote{\tt s.lee@kias.re.kr},
Zhao-Long Wang${}^{1,2}$\footnote{\tt zlwang@kias.re.kr},
and Piljin Yi${}^1$\footnote{\tt piljin@kias.re.kr}

\vskip 5mm
${}^1${\it School of Physics, Korea Institute
for Advanced Study, Seoul 130-722, Korea}\\[0.2cm]
${}^2${\it Institute of Modern Physics, Northwest University, Xian 710069, China}

\end{center}
\vfill

\begin{abstract}
\noindent
We count Higgs ``phase" BPS states of general non-Abelian quiver, possibly
with loops, by mapping the problem to its Abelian, or toric, counterpart
and imposing Weyl invariance later. Precise Higgs index computation
is particularly important for quivers with superpotentials; the Coulomb
``phase" index is recently shown to miss important BPS states, dubbed
intrinsic Higgs states or quiver invariants. We demonstrate how the
refined Higgs index is naturally decomposed to a sum over partitions
of the charge. We conjecture, and show in simple cases, that this decomposition
expresses the Higgs index as a sum over a set of partition-induced
Abelian quivers of the same total charge but generically of smaller rank.
Unlike the previous approach inspired by a similar decomposition of
the Coulomb index, our formulae compute the quiver invariants directly,
and thus offer a self-complete routine for counting BPS states.

\end{abstract}

\vfill
\end{titlepage}

\tableofcontents\newpage
\renewcommand{\thefootnote}{\#\arabic{footnote}}
\setcounter{footnote}{0}

\section{Quivers and Indices}

The low energy dynamics of BPS particles or BPS black holes in four dimensions
are most succinctly captured by quiver dynamics, which originate from wrapped
D-brane picture of such particles~\cite{Denef:2002ru,Denef:2007vg}
compactified on a Calabi-Yau 3-fold, where particle-like BPS states arise from D3-branes
wrapped on special Lagrangian 3-cycles. When the 3-cycle has topology of $S^3$,
the low energy dynamics of $n$ wrapped D3-branes would be $U(n)$ gauged quantum
mechanics with four supercharges. In the phase where the symmetry is broken to
$U(1)^{n}$, the triplet eigenvalues of the Cartan vector multiplets encode the
position of $n$ BPS particles along the noncompact $\mathbb{R}^3$, while the residual
Weyl group shuffles these $n$ identical particles. When more than one 3-cycles
are involved, each wrapped by D3-branes as well, we find additional chiral
multiplets, in bi-fundamentals, arising from open strings between each pair
of D3's. The number of such chiral fields is identified with the intersection number.

The quiver dynamics itself
can be further approximated by integrating out either vector multiplets or
chiral multiplets. The two such descriptions are called Higgs and Coulomb ``phase"
descriptions, respectively. The word ``phase" here is very misleading, although it
is used conventionally, as the quiver
dynamics in question is an one-dimensional system and thus the vacuum expectation values
do not imply superselection sectors. It merely refers to particular integrating out
procedure, which may or may not be reliable depending on the massgap, although
Supersymmetry tends to protect quantities like index further. When both sides are reliable,
 we expect the computed indices from the two sides to agree with each other.
This is the case, as far as we know, when the quiver has no loop \cite{Denef:2002ru,Sen:2011aa}.

Generally speaking, the Higgs description, better suited
for large Fayet-Iliopoulos (FI) constants,  $\theta$'s, is more reliable
as the massive vector multiplet fields being integrated out tend to have uniformly large
mass of order $m^2\sim \theta$, justifying the procedure.
The Coulomb description, suitable for small $\theta$'s, provides a more intuitive
picture of the wall-crossing via its multi-center picture; The positions of
charge centers are encoded in the Cartan part of the vector  multiplets.
While physically more appealing, this latter Coulomb description turns out to involve
various subtleties. Identification of the correct index theorem was rigorously argued
only very recently~\cite{Kim:2011sc}, and does not follow from naive truncation to classically flat
part of collective coordinates. The derivation has to invoke localization that breaks
the natural four supercharges of the low energy dynamics down to one. Another important
subtlety arises for cases with the so-called scaling regime, where one finds classical
multi-center solutions with mutual distances arbitrarily small.  For the latter class,
for which the relevant quiver dynamics must have at least one closed loop, the naive massgap
$m\sim 1/\theta$ for the chiral multiplets fails.

Nevertheless, in the absence of scaling regimes (thus, in the absence of loop),
the Coulomb description is very useful as it can be derived as the  BPS
soliton or black hole dynamics from the underlying $N=2$ $D=4$ theory, and it shows
a clear intuitive picture of wall-crossing phenomena as multi-particle bound state
physics~\cite{Lee:1998nv,Gauntlett:1999vc,Stern:2000ie,Denef:2000nb,Ritz:2000xa,Argyres:2001pv}.
This Coulomb description has been derived, ab initio, for Seiberg-Witten dyons \cite{Kim:2011sc,Lee:2011ph}, i.e.,
from the field theory itself as low energy dynamics of UV-incomplete solitons;
As long as we can ensure the individual constituent particles are actually
present in the spectrum, the low energy interaction among them are reliable
when we stay very close to the relevant marginal stability wall.
Regardless of how we view such multi-particle dynamics, a rather complete
derivation of the Coulomb index had emerged very recently
\cite{Kim:2011sc,Manschot:2010qz,Manschot:2011xc}, which was then shown \cite{Sen:2011aa} to
be equivalent among themselves and to the Kontsevich-Soibelman conjecture \cite{KS}.
We will briefly revisit this Coulomb index in section 4.

The natural index in $N=2$ $D=4$ field theory is the second helicity trace
\begin{equation}
\Omega(\gamma)=-\frac12\,{\rm tr}_\gamma (-1)^{2J_3}(2J_3)^2
\end{equation}
where trace is over the one-particle Hilbert space of the given charge $\gamma$,
and $J_3$ is the helicity operator.
For four-dimensional $N=2$ field theory, there is a natural equivariant
extension, called Protected Spin Character (PSC)~\cite{Gaiotto:2010be}
\begin{equation}
\Omega(\gamma;y)=-\frac12\,{\rm tr}_\gamma (-1)^{2J_3}(2J_3)^2y^{2J_3+2I_3}
\end{equation}
with $I_3$ belonging to $SU(2)_R$ symmetry.  When we factor out
the universal half-hypermultiplet factor in the BPS supermultiplets,
these reduce to the more familiar Witten-type indices
as
\begin{equation}
\Omega(\gamma)={\rm tr}'_\gamma (-1)^{2J_3}
\end{equation}
and
\begin{equation}
\Omega(\gamma;y)={\rm tr}'_\gamma (-1)^{2J_3}y^{2J_3+2I_3} \ .
\end{equation}
In the low energy description of these BPS objects, we effectively
compute the latter, after removing the free center-of-mass part of
the low energy dynamics.

In particular, for quiver dynamics,
PSC descends to \cite{Lee:2012naa,Lee:2012sc}
\begin{equation}
\Omega(\gamma;y)={\rm tr}'_\gamma (-1)^{2J_3}y^{2J_3+2I}
\end{equation}
where, as the quiver dynamics is a gauged quantum mechanics with
four supercharges, $SU(2)_J$ rotation generated by $J_a$'s is now
an R-symmetry of the quiver dynamics while $I$ generates the other
R-symmetry $U(1)_I$.
For Higgs ``phase," it has been argued that this equivariant
index is computed by (shifted) Hirzebruch characters,
\begin{eqnarray}
\Omega_{\rm Higgs}(y)= \sum_{p=0}^d\sum_{q=0}^d
(-1)^{p+q-d}y^{2p-d}{\rm dim} H^{(p,q)}\ ,
\end{eqnarray}
where $d$ is the complex dimension of the Higgs moduli space
of the quiver. The Higgs moduli space is always K\"ahler,
allowing us to use Hodge decomposition. This collapses to,
when $y=1$,
\begin{eqnarray}
\Omega_{\rm Higgs}= \sum_{n=0}^{2d} (-1)^{n-d}{\rm dim} H^{(n)}\ ,
\end{eqnarray}
which is the Euler number times $(-1)^d$. In terms of the
R-charges of the quiver, $2J_3\rightarrow (p+q)-d=n-d$ and $2I\rightarrow p-q$.

Actual computation of $\Omega_{\rm Higgs}$ is available for some
subfamilies of quivers. Reineke has given general formulae for the
Poincare polynomial of general quivers without loops~\cite{Reineke};
this can be thought of as Higgs counterpart of the Coulomb index
computations mentioned above. More interesting are $\Omega_{\rm Higgs}$
for quivers with loops, which neither of the above can address.
The equivariant index of an arbitrary Abelian cyclic quiver with generic
superpotential was computed in Refs.~\cite{Lee:2012naa}, and
along the way was found
a new class of BPS states \cite{Lee:2012sc,Lee:2012naa},
called intrinsic Higgs states. They were found to be wall-crossing-safe,
invisible from the Coulomb description, and of zero angular momenta. They are
typically far more numerous than Coulomb ``phase" states, given a quiver
with loops; These states are clearly important ingredients in understanding
microstates of $N=2$ single-center black holes, but they also appear in some
field theory BPS spectra, such as that of $N=2^*$ $SU(2)$ theory \cite{Alim:2011kw}.

A challenge we wish to face in this note is how to generalize these
Higgs index computations to general  {\it non-Abelian quivers
with superpotentials}.
For Abelian quivers that have been studied, the index is computable
relatively easily because  Higgs moduli spaces are  embedded
in  toric varieties. For non-Abelian cases, one encounters more general
symplectic quotients by non-Abelian groups and, with superpotentials,
has to intersect the zero loci of sections of vector bundles over such varieties.
A general procedure that can recast computation of indices on such
spaces to a problem in a bigger toric variety is known in the mathematical
literature~\cite{bkim1,bkim2,InterPairing}, which we will adapt to the
problem at hand. This effectively replaces any given non-Abelian quiver
by an Abelian one with the same total charge and of the same rank,
by splitting each non-Abelian node, say of rank $n$, to $n$ Abelian nodes.
Section 2 will declare the procedure and section 3 will elaborate with examples.

An interesting corollary of this Abelianization method is that the
end results have some similarity to the Coulomb ``phase" wall-crossing
formulae in Refs.~\cite{Manschot:2010qz,Kim:2011sc}. In the latter,
the gauge symmetry is spontaneously broken to the Cartan part,
with massless bosons encoding the positions of the particles.
The non-Abelian nature of the quiver enters only at the last step,
via the Weyl projection, which has been shown to result
in a sum over partitions of the charge \cite{Kim:2011sc}. Our Higgs
``phase" computation of index is very similar in spirit in that
we rely on Cartan subalgebra and the Weyl projection in the end.
This naturally leads us to suspect that our Abelianization procedure
parallels in some sense the index computation on the Coulomb side.
In section 4, we elaborate this idea further for simple examples,
and offer a conjecture on how the Higgs index can also be naturally
written as a sum over partitions of the charge in a manner that
parallels the Coulomb index partition sum.



After this work was completed, Ref.~\cite{Manschot:2013dua} appeared in the arXiv. 
There an intriguing transformation rule is suggested for the quiver invariants 
between different (non-Abelian) quivers related by mutation. Our formulae
provided in this note should be capable of verifing explicitly non-Abelian 
examples  in their work.

\section{How to Compute Higgs Index}\label{prescription}

As already explained in the previous section, Higgs phase index can be
computed as the Euler number $\chi(M)$ of the Higgs moduli space $M$, which, as we will shortly see, is constructed as
a complete intersection via F-terms, embedded in the D-term variety $X$.
As is well-known, with the aid of adjunction formula, certain invariants
of $M$, including its Euler number $\chi(M)$, are expressible in terms
of the ambient space data~\cite{GH}.
In case of Abelian quivers, the corresponding ambient space $X$
is a toric variety and hence, one can easily extract relevant invariants
in a straight-forward manner, by using simple combinatorial prescriptions
from toric geometry.
On the other hand, for general quivers with non-Abelian nodes,
it is more difficult to deal with the resulting D-term variety.

The upshot of the computational prescription for Higgs phase index is
to first ``Abelianize'' the quiver and to make use of the corresponding
``toric'' quiver variety $\tilde X$ as well as a complete intersection
$\tilde M$ therein.
One can then apply the usual toric techniques.
In this section, we shall briefly describe the index prescription
in full generality at the risk of making the presentation abstract;
some concrete, illustrative examples will follow in the ensuing section
for triangular quivers.

Before we proceed, it is important to note that we work in individual
branches of the quiver. In other words, we presume a definite choice of
FI parameters $\theta$. For each given branch, the Higgs
vacuum moduli space can be obtained via two steps; first, we perform
a symplectic reduction using D-term conditions, then, if a loop is
present, further impose F-term conditions. However, as was seen
in Ref.~\cite{Lee:2012sc} for Abelian quivers, we can make life slightly easier by
noticing that F-terms tend to simplify things. To make the long
story short, having a nontrivial F-term subvariety inside the D-term
variety often demands that some bi-fundamentals associated with certain
pairs of nodes should be set to zero. This is done to reduce the number
of F-terms, because F-terms tend to kill entire Higgs moduli space.
For each branch of the quiver, we end up freezing
certain sets of bi-fundamental fields to zero in order to obtain a
nontrivial $M$, which ``reduces" the quiver to be without loops
by removing links. See section 2 of Ref.~\cite{Lee:2012sc} or
Appendix~\ref{Appendix} here for an elaboration on this phenomenon.

\subsection{Abelianization and the Lift}\label{ss:ab}

Given a quiver ${\cal Q}$ with the gauge group $G=\prod_{v=1}^N U(r_v)$ and given
a choice of branch (or choice of FI constants), we have $X=\mu^{-1}_{G}(0)/G$
where $\mu_{G}$ is the moment map from the D-term, with
the shift by FI constants $\theta$ understood.
Then we obtain the true moduli space $M$ by further imposing F-term conditions.
The gauge group that actually participates in the quotient is $G/U(1)$ since
there is always one overall $U(1)$ that acts trivially on all chiral fields.
Again, the choice of branch imposes on us to set certain bi-fundamental fields
to zero identically, for otherwise $M$ is empty.

To such a non-Abelian quiver $\cQ$, we associate an Abelianized quiver
$\tilde{\cal Q}$, obtained by splitting each of the non-Abelian nodes of ${\cal Q}$,
say, of rank $r_v$, into $r_v$ Abelian nodes, and simply duplicating the arrows as well as the FI constants.
See Figs.~\ref{f:quiver1} and~\ref{f:quiver2} for an example.
Via this Abelianization, we reduce the gauge multiplets to those associated with the
Cartan subgroup $T=\prod_{v=1}^N U(1)^{r_v}$, but keep the same bi-fundamental
field contents. (Again, $T/U(1)$ acts nontrivially on the
chiral fields.) With such an Abelianized quiver $\tilde{\cal Q}$, we end up
in the territory of toric geometry. Keeping the same FI parameters, we
find the D-term induced variety, $\tilde X=\mu^{-1}_{T}(0)/T$,
and the subvariety $\tilde M$ obtained by imposing F-term conditions as well.
Thus, $\tilde M$ can be thought of as the Higgs moduli space of $\tilde{\cal Q}$
in the given branch. Finally, a useful intermediary that will eventually connect
the two D-term varieties $X$ and $\tilde X$ is the space
$$Y:=\mu^{-1}_{G} (0) / T \ , $$
which can be regarded as a bundle over $X$ and also a subvariety of $\tilde X$.

\begin{figure}[h]
\centering
\includegraphics[scale=0.55]{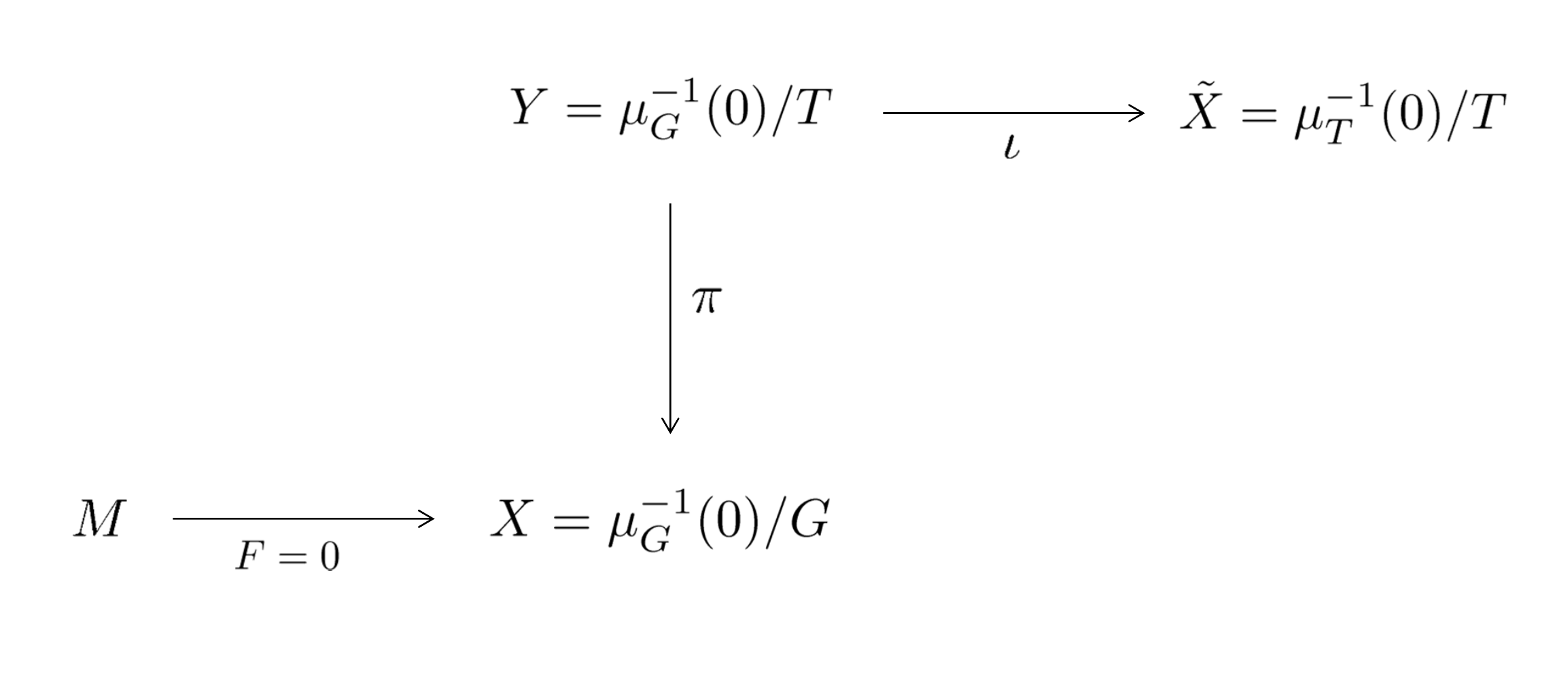}
\caption{\footnotesize Schematics of the Abelianization of the index computation.
The main point is that an index computation on $X$, or on its submanifold $M$,
can be lifted to a far more straightforward computation in the toric variety
$\tilde X$, obtained as the symplectic reduction by the Cartan subgroup $T$ of the gauge group $G$.
$\mu_H$ is the symplectic moment map associated with the group $H$ acting on the flat
K\"ahler space of chiral multiplet scalars. }\label{bundle}
\end{figure}

Refs.~\cite{InterPairing,Martin} lay down a simple procedure for
lifting topological invariants on $ X$ to $\tilde X$, thereby bridging the two spaces.
For any given cohomology class $a \in H^\star (X)$, the bridging rule
states that
\beq \label{interpairing}
\int_X a = \frac{1}{|W|} \int_{\tilde X} \hat a \wedge e(\bold \Delta)  \ ,
\eeq
where $W$ is the Weyl group of the gauge group $G$ for the non-Abelian quiver,
and $e(\bold \Delta)$ is the Euler class of $\bold \Delta$, the Whitney sum of line bundles associated
with the ``off-diagonal" part, $G/T$, of the gauge group, that is,
\beq\label{vander}
{\bold \Delta}\equiv \bigoplus_{\alpha\in\Delta} {\cal L}_\alpha\ .
\eeq
Note that $\Delta$ denotes the set of roots of $G$ while $\bold \Delta$
in bold denotes the corresponding vector bundle.
This bundle is naturally decomposed
as ${\bold\Delta}={\bold\Delta}_+\oplus{\bold\Delta}_-$ according to
the usual decomposition of $\Delta$ into the positive and the negative parts.
Here, ${\bold\Delta}_+$ is the (holomorphic) vector bundle that is tangent to
the fibre  of $\pi:Y\to X$. We will shortly see how to express $e({\bold \Delta})$
in terms of the toric data for $\tilde X$.

The nontrivial part of the bridging rule~\eqref{interpairing} is obviously the
lift of $a \in H^\star (X)$, denoted by $\hat a \in H^\star (\tilde X)$.
 Lift $\hat a$ is defined via the intermediary, $Y =\mu^{-1}_{G} (0) / {T}$, which naturally admits an inclusion
$\iota:Y \hookrightarrow \tilde X$ and a projection $\pi:Y\to X$,
in such a way that the relation
\begin{equation}\label{lift}
\pi^\star a = \iota^\star  \hat a \ ,
\end{equation}
holds on $Y$.
While this does not determine the lift $\hat a$ uniquely, given $a$,
whatever ambiguity there might be is killed by $e({\bold \Delta})$ that follows
on the right hand side of (\ref{interpairing}).
When the cohomology element $a$ is a multiplicative class $m$ associated with
the (holomorphic) tangent bundle $\cT X$, its lift turns out to be given as
\beq\label{TXDelta}
\widehat {m(\cT X)}=\frac{ m(\cT \tilde X)}{m({\bold\Delta_+})\wedge m({\bold\Delta_-})}
=\frac{ m(\cT \tilde X)}{m({\bold\Delta})}
\eeq
In the next section, we will see how this arises for
general quiver varieties in the course of evaluating index by
directly constructing a lifted bundle  $\widehat {\cT X}$ over
$\tilde X$ such that $\pi^\star \cT X = \iota^\star  \widehat {\cT X}$.

\subsection{Quivers without Loops, the Indices, and the Fans}

For quivers without loops, and thus, with no F-terms present, the above prescription
applies directly and simply since the Higgs moduli space is a symplectic
reduction, $X$, of the flat space of bi-fundamental chiral fields. When the quiver
has a loop, the superpotential will complicate the space further via
F-term constraints, which we will address in subsection \ref{2.4}.

The simplest invariant is the Euler number of $X$, $$\chi(X)=\sum_n (-1)^n \,b_n(X)\ , $$ that
counts the Higgs BPS states when the quiver in question has no
loops and thus no F-terms. In terms of the Chern class, $c$,
we find
\bea\label{Abel1}
\Omega [X]&=&(-1)^d\;\chi(X) \cr\cr
&=& (-1)^d\int_X c( \cT X) \cr\cr
&=&
\frac{(-1)^d}{|W|}\int_{\tilde X} \widehat{c(\cT X)}\wedge  e({\bold\Delta}) \cr\cr
&=&\frac{(-1)^d}{|W|}\int_{\tilde X} {c(\cT \tilde X)} \wedge  \frac{e({\bold\Delta})}{c({\bold\Delta})}  \ ,
\eea
where $d \equiv {\rm dim}_\IC (X)$ is the complex dimension of the Higgs moduli space.
The extra sign factor in front is there so as to count each hypermultiplet as $+1$.

A well-known equivariant version of the Euler number is the refined Euler
character, available upon the Hodge decomposition as
\beq
\chi_\xi = \sum\limits_{p \geq 0} \chi^p \, \xi^p \ , \quad \text{with}
\quad \chi^p = \sum\limits_{q\geq0} (-1)^q\,h^{p,q} \ ,
\eeq
which reduces to the Euler number when $\xi = -1$.
Recall that $\chi_\xi(X)$ is computed via the class (see for instance Ref.~\cite{Hirzebruch})
\beq
{\rm Td}(\cT X)\wedge {\rm ch}_\xi(\cT^* X)
\eeq
where ${\rm Td}$ and ${\rm ch}_\xi$ are the multiplicative classes associated, respectively, with $f_{\rm Td}(x)=x/(1-e^{-x})$ and $f_{\rm ch_\xi}(x)=1+\xi e^x$. The Abelianization asserts that this quantity can be computed as
\beq\label{HRR}
\chi_\xi(X)=\int_X {\rm Td}(\cT X) \wedge {\rm ch}_\xi (\cT^* X)=
\frac{1}{|W|}\int_{\tilde X} \frac{{\rm Td}(\cT \tilde X) \wedge
{\rm ch}_\xi (\cT^* \tilde X)}{{\rm Td}({\bold \Delta}) \wedge {\rm ch}_\xi
({\bold \Delta^*})}\wedge e({\bold\Delta})\ .
\eeq
We will show examples of these equivariant and non-equivariant indices
in the next section.

Alternatively, we can directly consider the topological class
associated with the refined Higgs index
\beq
\Omega(y)=(-y)^{-d}\;\chi_{\xi=-y^2}\ ,
\eeq
which will be more useful in interpreting the Abelianization physically in
section 4.
For this, it is convenient to view
the factor $(-y)^{-d}$ as a (trivial) multiplicative class associated with the
constant function $f_{\rm c}(x)=(-y)^{-1}$, whereby we find $\Omega(y)$ is directly
computed by another multiplicative class $ \omega_y$ associated with the
function
\bea
f(x)&=&f_{\rm c} (x) \cdot f_{\rm Td}(x) \cdot f_{\rm ch_\xi} (-x)\\
&=& \frac{x}{(1-e^{-x})}\cdot (ye^{-x}-y^{-1})\ ,
\eea
where $\xi$ has been replaced by $-y^2$.
In other words,
\beq
\Omega(y)[X]=\int_X \omega_y(\cT X)\ ,
\quad \omega_y(\cT X)\equiv \prod_\mu \left[x_\mu\cdot \left(\frac{ye^{-x_\mu}-y^{-1}}{1-e^{-x_\mu}}\right) \right] \ ,
\eeq
with eigen-forms $x_\mu$ of the  curvature of the holomorphic tangent bundle $\cT X$.
Again, we have
\beq\label{HRR2}
\Omega(y)[X]
=\frac{1}{|W|}\int_{\tilde X} {\omega_y (\cT \tilde X)}
\wedge \frac{e({\bold\Delta})}{\omega_y({\bold \Delta})} \ ,
\eeq
in the lifted form.

To understand the origin of the contribution from $\bold \Delta$ in Eq.~\eqref{TXDelta}
(and consequently in Eqs.~\eqref{Abel1},~\eqref{HRR} and~\eqref{HRR2}), it is useful
to consider how the lift $\widehat {\cT X}$ of the tangent bundle ${\cT X}$ is
related to the tangent bundle ${\cT \tilde X}$ of the Abelianized variety $\tilde X$.
The relevant exact sequence for general quiver can be written as \beq\label{eq:geneuler}
0 \:\:\to\:\: \left[\bigoplus_{i=1}^r\mathcal {\cal O}_i \right]\bigoplus
\left[\bigoplus_{\alpha\in \Delta}{\cal L}_\alpha\right]
\:\:\to\:\:\bigoplus_{\rho \in \Sigma^{(1)}} \mathcal L_\rho\:\:\to\:\:
\widehat {\cT X} \:\:\to\:\:    0 \ ,
\eeq
with $r= {\rm rk}\; G - 1$, where ${\cal O}_i$'s are $r$ copies
of the trivial line bundle, call it ${\cal O}$, over $\tilde X$.
The label $i$  serves as a reminder how the corresponding Cartan generator determines
the map ${\cal O}_i \;\;\to\;\; \bigoplus {\cal L}_\rho$. In turn, the latter is a
sum of line bundles, ${\cal L}_\rho$, where $\rho$ belongs to the collection
of one-dimensional cones, $\Sigma^{(1)}$, in the ``fan,'' $\Sigma$, for the Abelianized toric
variety $\tilde X$.

For any multiplicative class $m$, then, we have
\beq
\widehat{m(\cT X)}=m(\widehat{\cT X})=
\frac{\prod_{\rho\in \Sigma^{(1)}} m({\mathcal L_\rho})}{ [ m({\cal O})]^r\wedge
\prod_{\alpha\in \Delta}m({\cal L}_\alpha)}
=\frac{\prod_{\rho\in \Sigma^{(1)}} m({\mathcal L_\rho})}{[ m({\cal O})]^r\wedge
 m({\bold\Delta}) } \ .
\eeq
As the Euler sequence for the Abelianized D-term variety $\tilde X$ is given by
\beq\label{eq:geneuler_ab}
0 \:\:\to\:\: \left[\bigoplus_{i=1}^r\mathcal {\cal O}_i \right]\:\:\to\:\:
\bigoplus_{\rho \in \Sigma^{(1)}} \mathcal L_\rho\:\:\to\:\:
{\cT \tilde X} \:\:\to\:\:   0 \ ,
\eeq
we conclude that
\beq
\widehat {m(\cT X)}
=\frac{ m(\cT \tilde X)}{m({\bold\Delta})}\ , \quad
m(\cT \tilde X)=\frac{\prod_{\rho\in \Sigma^{(1)}} m({\mathcal L_\rho})}{[ m({\cal O})]^r} \ .
\eeq
Given the toric data for $\tilde {\cal Q}$, this carries all the information
required to express the (equivariant) index as the integral of a specific cohomology class over $\tilde X$.

To evaluate this integral, we still need to determine the various intersection numbers, which can only be understood through the complete fan structure for $\tilde X$.
First of all, associated with the Abelianized quiver is a
corresponding ``charge matrix'', denoted by $Q=[Q_{ve}]$, each row of which lists the
charges of the bi-fundamental fields under each $U(1)$ gauge group.
Note that the row and the column indices for the matrix range over the regions $1\leq v \leq {\rm rk}(T/U(1))\equiv r$
and $1\leq e \leq |\Sigma^{(1)}| \equiv k$, respectively.
The charge matrix itself has a certain amount of information
on the toric variety $\tilde X$.
For instance, any multiplicative class of the tangent bundle $\cT \tilde X$,
say, the Chern class of $\cT \tilde X$, can be expressed explicitly in terms of $Q$,
\beq
c(\cT\tilde X) = \prod\limits_{e=1}^k [1+ \sum\limits_{v=1}^r Q_{ve} J_v] ,
\eeq
where $J_{v=1,\cdots, r}$ form a basis of the $H^2(\tilde X)$, which turns out to be of rank $r$.
However, the charge matrix does not uniquely determine the fan; in toric terms,
its rows correspond to the linear relations of the rays in the fan,
but the incidence information for higher-dimensional cones is missing.

For the rest of this subsection, we summarize how the complete fan structure for the toric variety $\tilde X$ is determined from the given quiver data, and also present the recipe for the intersection numbers.
The technical details will not be needed in reading the rest of this paper and the way we state the procedures here is by no means pedagogical.
Interested readers are kindly referred to the excellent texts~\cite{fulton, oda, cox:review, coxnkatz} for a more complete review.

It turns out that the charge matrix, when equipped with $\theta$ values
assigned to the quiver nodes (that is, a choice of $\theta$-stability
criterion), does determine the fan completely; the notion of stability
of reduced quivers can be defined accordingly, from which the fan structure is determined~\cite{hille}.
Practically, however, the procedure illustrated in Ref.~\cite{iritani}
can be more accessible, which goes as follows:
an index set $\mathcal A$ is defined as
\beq
\mathcal A =\{I\subset \Sigma^{(1)}\;|\;  \exists \,a_e>0~\text{such that~}
\theta_v = \sum\limits_{e \in I} Q_{ve}\; a_e \text{ , for } 1\leq v \leq r  \} \ ,
\eeq
and by collecting the maximal elements of the complement $\mathcal A^c$ inside
the power set of $\Sigma^{(1)}$, one obtains the Stanley-Reisner ideal $\mathcal I_{\rm SR}$,
from which the corresponding fan $\Sigma$ is constructed as\footnote{By abuse of notation we denote the cone $\sigma={\rm Span}(I) \subset \IR^d$ simply by the set, $I$, of its generators. }
\beq
\Sigma=\{I\subset \Sigma^{(1)}\;|\; I \not\supset S \text{ , for all } S \in \mathcal I_{\rm SR}  \} \ .
\eeq

Now given the charge matrix $Q$ and the fan $\Sigma$ for the toric variety $\tilde X$, the intersection
numbers defined as
\beq
\kappa_{v_1v_2\cdots v_d} \equiv \int J_{v_1}\wedge \cdots \wedge J_{v_d}\ ,  \quad \text{with}\quad 1\leq v_{s=1, \dots, d} \leq r \ ,
\eeq
are determined by simultaneously solving the linear equations of the following form,
\beq
\sum\limits_{v_1=1}^r \cdots \sum\limits_{v_d=1}^{r}\kappa_{v_1 v_2 \cdots v_d}\,
Q_{v_1e_1} \, Q_{v_2e_2}\, \cdots \,Q_{v_de_d}=  \left\{ \begin{array}{rl}
 1 &\mbox{ if $\{\rho_{e_1}, \cdots, \rho_{e_d}\} \in \Sigma^{(d)}$ } \ ,  \\
 0 &\mbox{ if $\{\rho_{e_1}, \cdots, \rho_{e_s}\} \notin \Sigma^{(s)}$ with $s \leq d$ } \ ,
       \end{array} \right.
\eeq
where $\rho_e$ denote the ray corresponding to the $e$-th column of $Q$, and $\Sigma^{(s)} \subset \Sigma$,  the collection of $s$-dimensional cones for $1\leq s \leq d$.

\subsection{An Illustration: Grassmannian $X$}

As an illustration, let us consider the quiver with gauge group $G=U(r)\times U(1)$, the two nodes for which are linked by $\kappa$ arrows. The Higgs moduli space $X=Gr(r,\kappa)$ is
the Grassmannian, whose indices are of course well-known already.
Nevertheless, let us proceed
to compute its topological invariants following the Abelianization
procedure. The Abelianized variety $\tilde X= \left(\IP^{\kappa-1}\right)^{r}$
consists of $r$ copies of projective spaces and we denote by $J_{i=1,\dots,r}$
the K\"ahler class of each copy.
The intersection structure is simple;
\beq
\int_{\tilde X} (J_1)^{\kappa-1}\wedge (J_2)^{\kappa-1}\wedge \cdots\wedge (J_{r})^{\kappa-1}=1 \ ,
\eeq
is the only nonvanishing intersection number.

From the Euler sequence~\eqref{eq:geneuler_ab} for $\tilde X$, or that for each of the projective spaces
\beq
0 \:\:\to\:\: \cO_{\IP^{\kappa-1}} \:\:\to\:\: \cO_{\IP^{\kappa-1}}(1)^{\oplus \kappa} \:\:\to\:\: \cT\IP^{\kappa-1} \:\:\to\:\: 0 \ ,
\eeq
we find
\bea
c(\cT \tilde X)&=& \prod_i(1+J_i)^{\kappa} \ , \cr
{\rm Td} (\cT \tilde X) &=&\prod_i\left(\frac{J_i}{1-e^{-J_i}}\right)^\kappa \ ,\cr\cr
{\rm ch}_\xi(\cT^*\tilde X) &=& \prod_i\frac{(1+\xi e^{-J_i})^\kappa}{(1+\xi)}\ ,\cr\cr
\omega_y (\cT \tilde X) &=& \prod_i \left[ \left(\frac{J_i}{1-e^{-J_i}} \right)^\kappa
\cdot \frac{(ye^{-J_i}-y^{-1})^\kappa}{y-y^{-1}}\right] \ ,
\eea
where the products run over the range $1\leq i \leq r$.
The factors associated with ${\bold\Delta}$ can be read off from the
off-diagonal parts of $U(r)=G/U(1)$. The $J_i$'s are associated with $U(1)^{r}=T/U(1)$,
under which the off-diagonal parts are labeled by a pair of ordered indices, $i\neq j$,
which have the charge of the form
$$(0,\cdots, 0,1,0,\cdots, 0,-1,0, \cdots,0) \ .$$
So,
$c_1({\cal L}_{\alpha=(ij)})=J_i-J_j$, which leads to
\bea
c({\bold \Delta}) &=& \prod_{i\neq j}(1+J_i-J_j)\ ,\cr
{\rm Td} ({\bold\Delta}) &=&\prod_{i\neq j}\frac{J_i-J_j}{1-e^{-J_i+J_j}}\  ,\cr \cr
{\rm ch}_\xi({\bold\Delta^*}) &=& \prod_{i\neq j}(1+\xi e^{-J_i+J_j}) \ ,\cr\cr
\omega_y ({\bold\Delta}) &=&\prod_{i\neq j}\left[\left(J_i-J_j\right)
\cdot\left(  \frac{ye^{-J_i+J_j}-y^{-1}}{1-e^{-J_i+J_j}}            \right)\right]  \ ,
\eea
and
\bea
 e({\bold\Delta}) &=&\prod_{i\neq j} (J_i-J_j) \ .
\eea
These combined, Eq.~\eqref{Abel1} (and Eq.~\eqref{HRR2}, respectively) reproduces the (refined) Higgs index of the Grassmannian faithfully. We will come back to this example in
section 4, and  try to give the resulting index formula a little more
physical interpretation.

\subsection{Loops, the Superpotential, and the Normal Bundle}\label{2.4}

Computation of any multiplicative class for $M$ embedded in $X$ is
straightforward as long as we understand the normal bundle ${\cal N}$
of this embedding.
The general rule states that
\beq
\int_M m(\cT M) =\int_X {m(\cT X) }\wedge \frac{e(\cN )}{m({\cal N} )}\ .
\eeq
For the problem at hand, we are interested in $m=c$ for the unrefined index, and in either
$m={\rm Td}\wedge {\rm ch}_\xi^*$ or
$m=\omega_y$ for the refined one.\footnote{The ${\rm ch}_\xi^*$
class of a bundle denotes the ${\rm ch}_\xi$ class of the dual bundle.}
For Abelian quivers this general formula
has been used very fruitfully in Refs.~\cite{Lee:2012sc,Lee:2012naa,Bena:2012hf}, where
a new class of BPS states, intrinsic Higgs states, was discovered.
For non-Abelian quivers, this is again lifted to the Abelianized
form,
\beq
\int_M m(\cT M) 
= \frac{1}{|W|}\int_{\tilde X}
{{m(\cT \tilde X)}}\wedge \frac{\widehat{e(\cN )}}{\widehat{m({\cal N} )}}\wedge \frac{e({\bold\Delta})}{m({\bold\Delta})}\ ,
\eeq
so it remains to understand how the normal bundle $\cN$ of $M$ in $X$ is lifted
to a bundle $\widehat \cN$ over $\tilde M$ in $\tilde X$.

For quivers with a loop, $\cQ$, the superpotential $\cW$ generates a F-term
constraint
\beq \partial \cW =0  \eeq
for each chiral multiplet in the quiver and defines the embedding of $M$ in $X$.
Note that, with a generic choice
of superpotential and a generic choice of FI constants, the D-term and
the F-term constraints are independent. When we Abelianize $\cQ$ to
$\tilde \cQ$, we are removing non-Cartan part of the
D-term constraints from the data but leave the chiral field contents
and the superpotential thereof intact. This shows that, generically the
fibre of $\cN$ coincides with that of $\tilde\cN$, i.e., the normal
bundle of the Abelianized Higgs moduli space $\tilde M$ embedded into
its D-term ambient $\tilde X$. This is in contrast with how fibre of
$\cT \tilde X$ is a sum of the fibre of $\cT X$ and that of ${\bold \Delta}$.

In fact, a natural and simple lift of the normal bundle and its
topological classes dictates
\beq
\widehat{m(\cN)}\; = \; m(\tilde \cN)\ ,\qquad \widehat{e{(\cN)}}\; = \; e(\tilde \cN)
\eeq
In other words,
\beq\label{looped}
\frac{1}{|W|}\int_{\tilde X} {m(\cT \tilde X)}\wedge
\frac{e(\tilde \cN)}{m({\tilde \cN})}\wedge \frac{e({\bold\Delta})}{m({\bold\Delta})}\ ,
\eeq
computes the $M$-integral of the multiplicative class $m$ via its
Abelianized D-term variety $\tilde X$ and the embedded $\tilde M$.
Again, we will be working with toric varieties, so all quantities here
can be straightforwardly read-off from $\tilde \cQ$.

\section{Examples with a Loop: Triangular Quivers}

Having seen the general prescription for computing the Higgs phase index,
we shall illustrate it, in this section, with simplest examples with
an oriented loop: the triangular quivers.
\subsection{A Simplest Non-Abelian Triangular Quiver}\label{simple_ex}
\begin {figure}[h]
\centering
\includegraphics[scale=0.3]{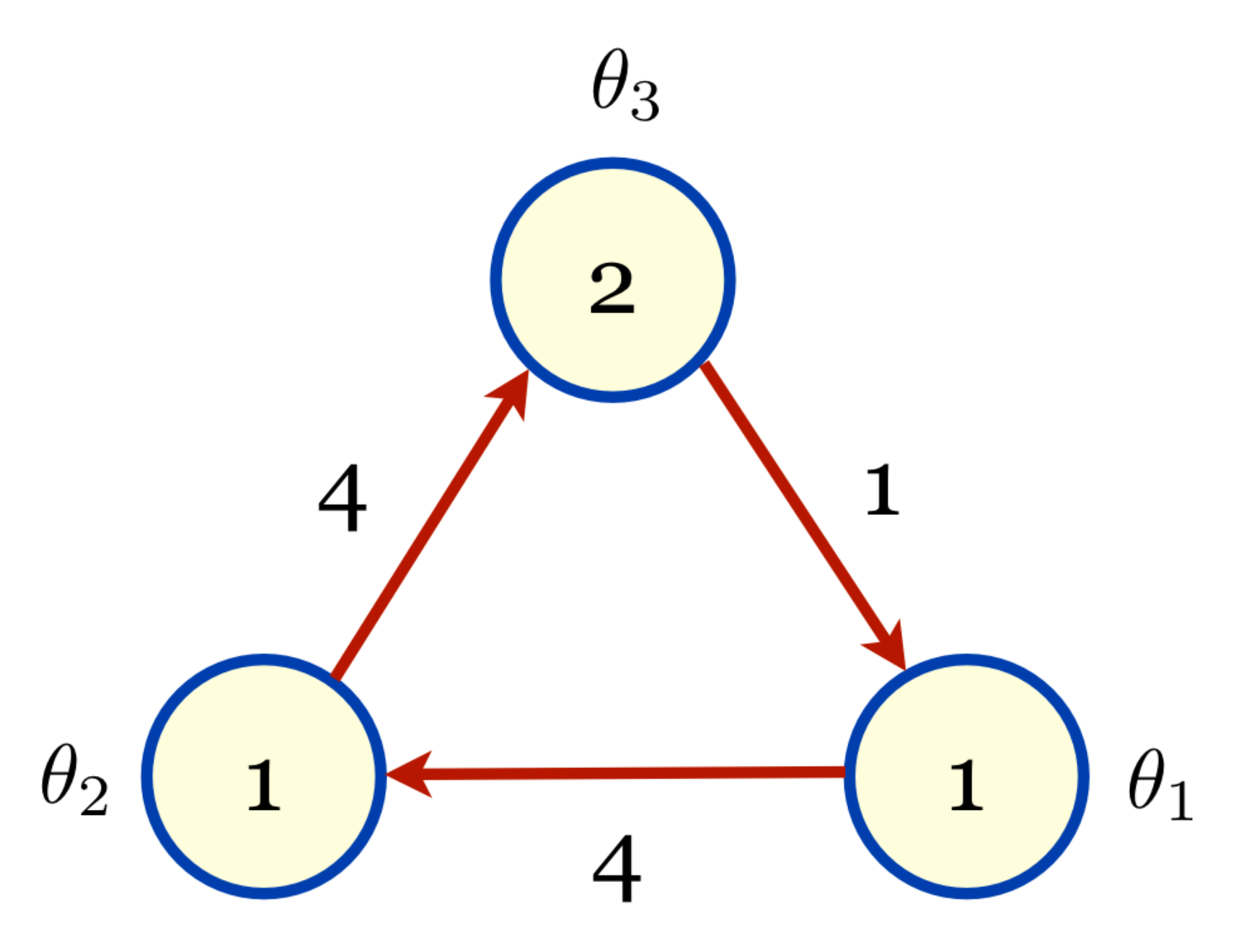}
\caption{\small Non-Abelian triangular quiver with adjacency matrix~\eqref{adj_quiver1}.}\label{f:quiver1}
\end{figure}

Let us consider the quiver with the adjacency matrix \\
\beq\label{adj_quiver1}
A={\left(\ba{ccc} 0 & 4 & -1 \\ -4 & 0 & 4\\ 1 & -4 & 0 \ea\right)} \ ,
\eeq
and the dimension vector $\bold d=(r_1,r_2, r_3)=(1,1,2)$, as depicted in Fig.~\ref{f:quiver1}.
The computation of $\chi(M)$ is illustrated in the branch where $\theta_1 <0$ and $\theta_3 >0$ so that the single bi-fundamental field from node $3$ to node $1$ gets a zero VEV.
Firstly, by imposing D-terms, one is led to the ambient variety $X=\IP^3 \times {Gr}(2,4)$.
The vacuum moduli space $M$ is then embedded in $X$ through the F-term,
defined as a section of the rank-2 vector bundle associated with
the vanishing $(\bar{\bold 2}, { \bold  1})$-bi-fundamental field under $U(2)_3\times U(1)_1$,
where the subscripts for gauge groups label the nodes.

\begin {figure}[h]
\centering
\includegraphics[scale=0.37]{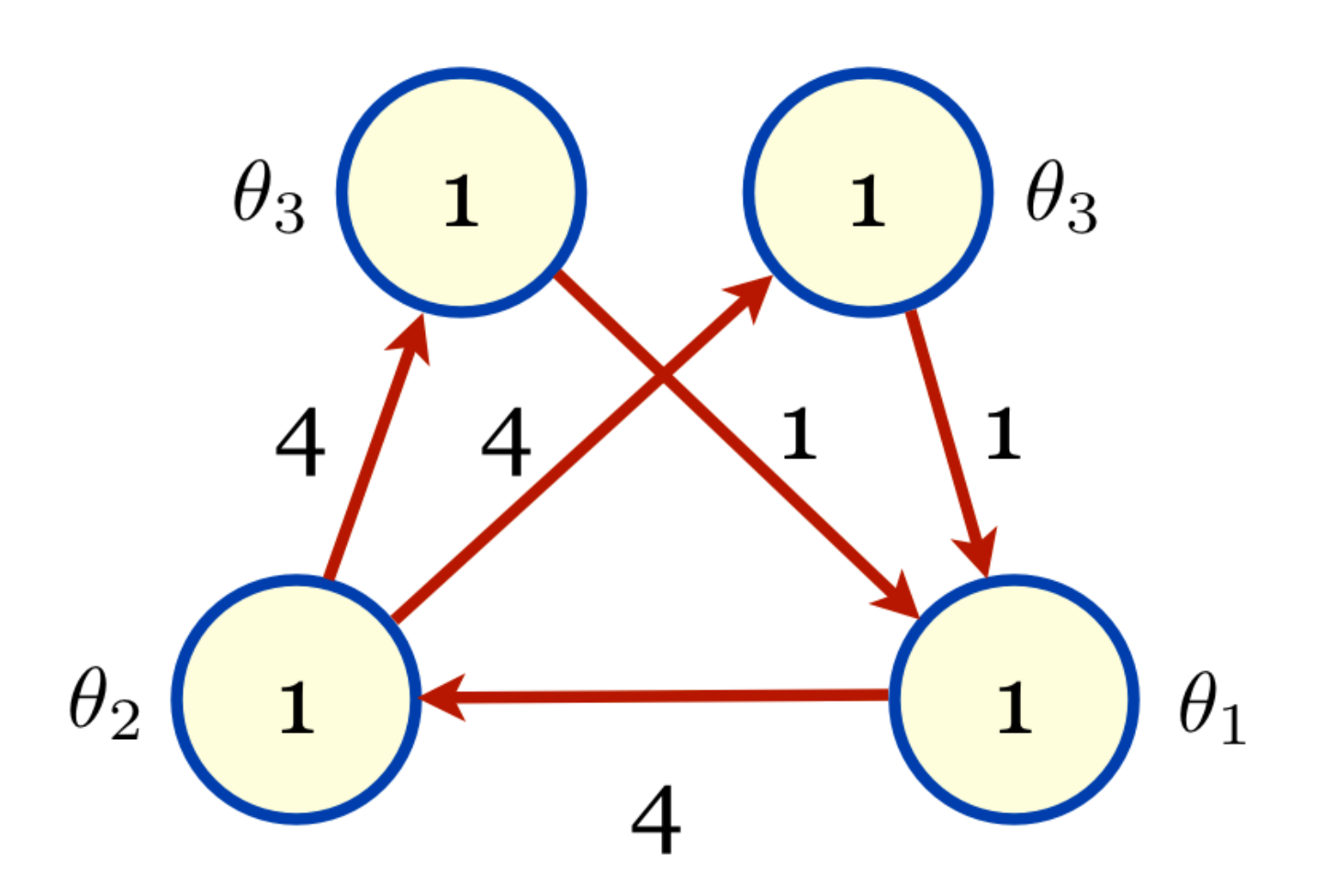}
\caption{\small Abelian quiver with adjacency matrix~\eqref{adj_quiver2},
obtained through Abelianization of the quiver in Fig.~\ref{f:quiver1}.}\label{f:quiver2}
\end{figure}

Now, we shall apply the general prescription of section~\ref{prescription}
and move towards the territory of line bundles on toric geometry,
as opposed to that of vector bundles on Grassmannian geometry.
Upon Abelianizing the quiver, the rank-2 node gives rise to
two Abelian nodes.
Thus, the resulting quiver is described by the following adjacency
 matrix
\beq\label{adj_quiver2}
{A}={\left(\ba{cccc} 0 & 4 & -1 & -1 \\ -4 & 0 & 4 & 4\\ 1 & -4 & 0 & 0 \\ 1 & -4 & 0 &0 \ea\right)} \ ,
\eeq
with the dimension vector ${\bold d}= (1,1,1,1)$, as depicted in Fig.~\ref{f:quiver2}.
It is easy to see that the corresponding D-term variety is
$\tilde X = \IP^3 \times \IP^3 \times \IP^3$. Note that the $Gr(2,4)$ piece of the original D-term variety $X=\IP^3 \times Gr(2,4)$ has led to the last two $\IP^3 \simeq Gr(1,4)$ factors of $\tilde X$ upon Abelianization.

By taking the multiplicative class $m$ in Eq.~\eqref{looped}
to be the Chern class $c$, one is thus led to the following expression
for $\chi(M)$ as an integral over $\tilde X = \IP^3_J \times \IP^3_{K_1} \times \IP^3_{K_2}$,
\bea
\chi(M)
&=&\frac{1}{2}\int_{\tilde X} {c(\cT \tilde X)}\wedge\frac
{e(\tilde \cN)}{c({\tilde \cN})}\wedge \frac{e({\bold\Delta})}{c({\bold\Delta})} \cr\cr  \label{form}
&=& \frac{1}{2}\int_{\IP_J^3\times\IP_{K_1}^3\times\IP_{K_2}^3}
\underbrace{ {(1+J)^4\wedge (1+K_1)^4 \wedge (1+K_2)^4} }_{c(\cT \tilde X) } \\
&&~~~~\wedge \underbrace{\frac{(J+K_1)\wedge(J+K_2)}{(1+J+K_1)
\wedge(1+J+K_2)}}_{e(\tilde \cN) \, \wedge \, c(\tilde \cN)^{-1}}
\wedge  \underbrace{\frac{(K_1-K_2)\wedge (K_2-K_1)}{(1+K_1-K_2)
\wedge (1+K_2-K_1)}}_{e(\bold \Delta)\,\wedge \,c(\bold \Delta)^{-1}}\cr\cr \label{contour}
&=&\frac12 \oint {\rm d}J \; {\rm d}K_1 \;{\rm d}K_2
\left(\frac{1+J}J\right)^4 \left(\frac{1+K_1}{K_1}\right)^4\left(\frac{1+K_2}{K_2}\right)^4 \\ \nn
&&~~~~\cdot \frac{(J+K_1)(J+K_2)}{(1+J+K_1) (1+J+K_2)}
\cdot \frac{(K_1-K_2)(K_2-K_1)}{(1+K_1-K_2)(1+K_2-K_1)} \ ,
\eea
where $J$, $K_1$ and $K_2$ denote the K\"ahler classes of the three
$\IP^3$ factors, respectively, and in the last step, via the trivial
intersection structure of $\IP^3$, the integration of the cohomology
class has switched to a contour integral around the origin.\footnote{Note that we could have had $-J$ replacing $J$ in Eq.~\eqref{form} to conform with the convention used in subsection~\ref{subsec3.2.} for a general triangular quiver. Under such a choice, $J$ would not lie in the K\"ahler cone and Eq.~\eqref{contour} should get an extra sign factor due to the negative intersection. }
Note that the $(2\pi i)^{-1}$ factor is implicit in each of the contour integral measures.
It is straightforward to evaluate Eq.~\eqref{contour}
and we obtain $\chi(M)=12$. 

As for the computation of the refined Euler character, we again apply
Eq.~\eqref{looped}, now with $m={\rm Td}\wedge {\rm ch}_\xi^*$,
\beq\label{chirefex}
\chi_\xi (M)= \frac{1}{2}\int_{\tilde X} {{\rm Td}(\cT \tilde X)}\wedge{{\rm ch}_\xi (\cT^* \tilde X)}\wedge
\frac{e(\tilde \cN)}{{\rm Td}({\tilde \cN})\wedge{\rm ch}_\xi({\tilde \cN^*})}
\wedge \frac{e({\bold\Delta})}{{\rm Td}({\bold\Delta})\wedge{{\rm ch}_\xi({\bold\Delta^*})} }
\eeq
where the four factors in the integrand are written in turn as
\bea \nn
{{\rm Td}(\cT \tilde X)}&=& \left(\frac{J}{1-e^{-J}}\right)^4
\left(\frac{K_1}{1-e^{-K_1}}\right)^4 \left(\frac{K_2}{1-e^{-K_2}}\right)^4 \cr\cr \nn
{{\rm ch}_\xi(\cT^* \tilde X)}&=&\frac{1}{(1+\xi)^3}\,
(1+\xi e^{-J})^4 (1+\xi e^{-K_1})^4 (1+\xi e^{-K_2})^4 \ , \\ \nn
\frac{e(\tilde \cN)}{{\rm Td}({\tilde \cN})\wedge{\rm ch}_\xi({\tilde \cN^*})}
&=& \frac{(1-e^{-J-K_1})(1-e^{-J-K_2})}{ (1+\xi e^{-J-K_1})(1+\xi e^{-J-K_2})} \ ,  \\ \nn
\frac{e(\bold \Delta)}{{\rm Td}(\bold \Delta) \wedge{{\rm ch}_\xi(\bold \Delta^*)}  }
&=& \frac{(1-e^{K_1-K_2})(1-e^{K_2-K_1})}{(1+\xi e^{K_1 - K_2})(1+\xi e^{K_2 - K_1})}  \ .
\eea
Similarly to the unrefined case, we are led to a straightforward contour integral and thereby obtain
\beq \label{ans}
\chi_\xi (M) = 1- 2 \xi + 3 \xi^2 -3 \xi^3 + 2 \xi^4 - \xi^5 \ ,
\eeq
which gives the refined Higgs index
\bea
\Omega(y)[M]&=&(-y)^{-d} \, \chi_{\xi=-y^2} (M) \\
&=& -\frac{1}{y^5} - \frac{2}{y^3} -\frac{3}{y} -  3y -2y^3 -y^5 \ .
\eea
As desired, for $\xi=-1$, the refined Euler character~\eqref{ans} does reduce to the Euler number $\chi(M)=12$.

\subsection{General Triangular Quivers}\label{subsec3.2.}
\begin {figure}[h]
\centering
\includegraphics[scale=0.3]{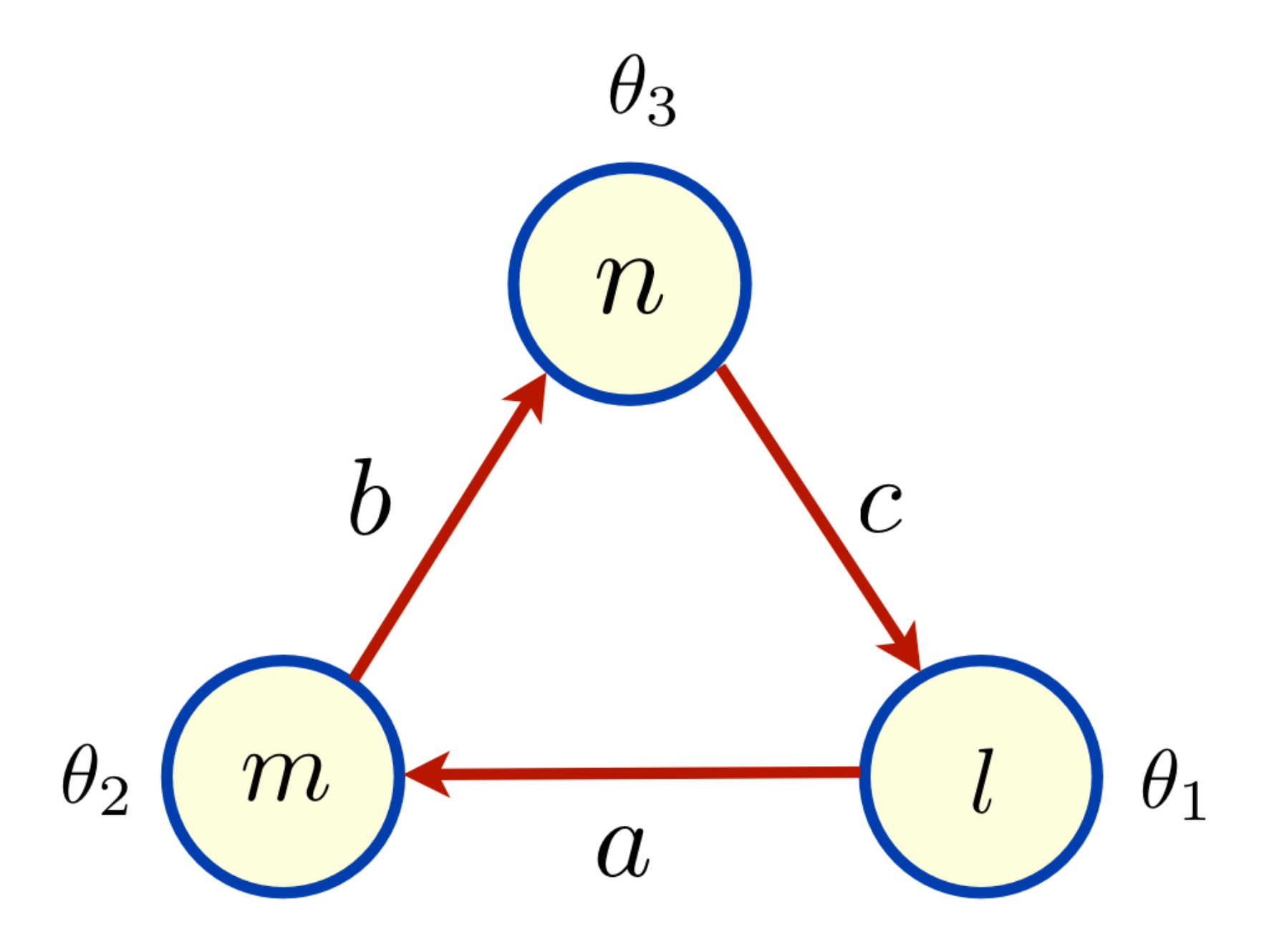}
\caption{\small A general non-Abelian triangular quiver}\label{f:general}
\end{figure}

Let us now consider triangular quivers in full generality
(see Fig.~\ref{f:general}).
In the previous example, the Abelian ambient variety $\tilde X$ was
a product of projective spaces and hence, the integration~\eqref{form} of a cohomology
class, for instance, turned into the contour integral~\eqref{contour} in a trivial manner.
In general, complications may arise due to the non-trivial intersection structure of $\tilde X$.
We are still in the territory of toric geometry, however, and
topological invariants can be obtained by some simple combinatorics.
Let us work in the branch where the $c$ fields that transform
as $(\bar{\bold n}, {\bold l})$ under $U(n)_3 \times U(l)_1$ vanish simultaneously.

\begin {figure}[h]
\centering
\includegraphics[scale=0.3]{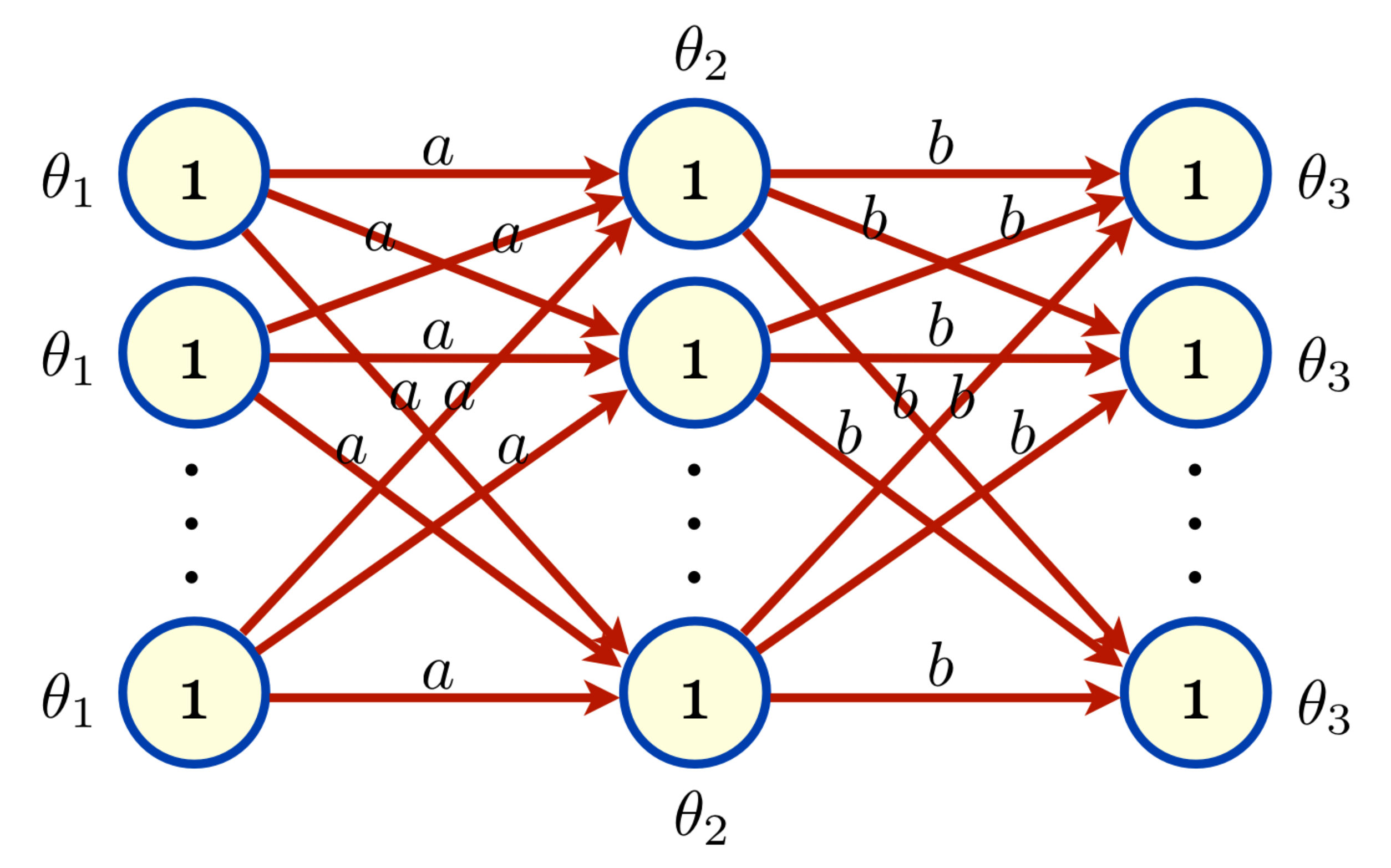}
\caption{\small Abelianization of the quiver in Fig.~\ref{f:general}; the $c$ vanishing fields are ignored here, given the branch in question.}\label{f:general_ab}
\end{figure}

Fig.~\ref{f:general_ab} depicts the Abelianization of the quiver in Fig.~\ref{f:general}.
Note that the $c$ vanishing fields have been ignored for the simplicity of drawing.
%
The D-term ambient space $\tilde X$ is a toric quiver variety and hence,
can be completely described by its fan, which itself is determined by
the charge matrix $Q$ together with $\theta$ values on the nodes (or the $\theta$-stability criterion).
Amongst the $l+m+n$ Abelian groups, that is, $U(1)_{A, i}$ for $i \in [1, l], [1, m], [1,n]$,
respectively, for $A=1, 2, 3$, one can ignore an overall $U(1)$ and we choose to take
\beq T/U(1)=\prod\limits_{i=1}^l U(1)_{1,i} \prod\limits_{j=1}^m U(1)_{2,j} \prod\limits_{k=1}^{n-1} U(1)_{3,k}\ ,
\eeq
with the last Abelian factor $U(1)_{3,n}$ quotiented from $T$.

Then, the $(l+m+n-1) \times (aml + bmn)$ matrix $Q$, in an appropriate arrow ordering, can be written as follows:
{\renewcommand{\arraystretch}{1.5}
\setlength\arraycolsep{3.5pt}
$$\tiny\begin{array}{l}
\left( {\begin{array}{c||cccccccccc|ccccccccc}
{\text {Node}}&\multicolumn{3}{c}{\overbrace{\quad\quad\quad\quad\quad\quad\quad}^{m~\text{columns}}}
&\multicolumn{3}{c}{\overbrace{\quad\quad\quad\quad\quad\quad\quad}^{m~\text{columns}}}
& \cdots & \multicolumn{3}{c}{\overbrace{\quad\quad\quad\quad\quad\quad\quad}^{m~\text{columns}}}
&\multicolumn{4}{c}{\overbrace{\quad\quad\quad\quad\quad\quad\quad\quad\quad\quad}^{n~\text{columns}}}
& \cdots & \multicolumn{4}{c}{\overbrace{\quad\quad\quad\quad\quad\quad\quad\quad\quad\quad}^{n~\text{columns}}} \\ \hline\hline
{{U(1)_{1,1}}}&{{-\bold 1_a}}&{{\cdots}}&{-\bold 1_a}&{{\bold 0_a}}&{{\cdots}}&{{\bold 0_a}}
& \cdots & \bold 0_a & \cdots & \bold 0_a & \bold 0_b & \cdots & \bold 0_b&\bold 0_b &\cdots
&\bold 0_b &\cdots & \bold 0_b& \bold 0_b \\

U(1)_{1,2}&{ \bold 0_a}&{ \cdots}&{ \bold 0_a}&{-\bold 1_a}&{\cdots }&{-\bold 1_a}
&\cdots &\bold 0_a&\cdots&\bold 0_a&\bold 0_b&\cdots&\bold 0_b&\bold 0_b &\cdots
& \bold 0_b&\cdots&\bold 0_b& \bold 0_b\\

\vdots&\vdots&\cdots&\vdots&\vdots&\cdots&\vdots &\cdots&\vdots&\cdots&\vdots&\vdots
&\cdots&\vdots&\vdots&\cdots&\vdots&\cdots&\vdots&\vdots\\

U(1)_{1,l}&\bold 0_a&\cdots&\bold 0_a&\bold 0_a&\cdots&\bold 0_a &\cdots&-\bold 1_a
&\cdots&-\bold 1_a&\bold 0_b&\cdots&\bold 0_b&\bold 0_b&\cdots& \bold 0_b&\cdots&\bold 0_b& \bold 0_b\\ \hline

U(1)_{2,1}&\bold 1_a&\cdots&\bold 0_a&\bold 1_a&\cdots&\bold 0_a &\cdots&\bold 1_a
&\cdots&\bold 0_a&-\bold 1_b&\cdots&-\bold 1_b&-\bold 1_b&\cdots&\bold 0_b&\cdots&\bold 0_b& \bold 0_b\\

\vdots&\vdots&\cdots&\vdots&\vdots&\cdots&\vdots &\cdots&\vdots&\cdots&\vdots&\vdots
&\cdots&\vdots&\vdots&\cdots&\vdots&\cdots&\vdots&\vdots\\

U(1)_{2,m}&\bold 0_a&\cdots&\bold 1_a&\bold 0_a&\cdots&\bold 1_a &\cdots&\bold 0_a
&\cdots&\bold 1_a&\bold 0_b&\cdots&\bold 0_b&\bold 0_b&\cdots&-\bold 1_b&\cdots&-\bold 1_b & -\bold 1_b\\ \hline

U(1)_{3,1}&\bold 0_a&\cdots&\bold 0_a&\bold 0_a&\cdots&\bold 0_a &\cdots&\bold 0_a
&\cdots&\bold 0_a&\bold 1_b&\cdots&\bold 0_b&\bold 0_b &\cdots&\bold 1_b&\cdots&\bold 0_b &\bold 0_b\\

\vdots&\vdots&\cdots&\vdots&\vdots&\cdots&\vdots &\cdots&\vdots&\cdots&\vdots&\vdots
&\cdots&\vdots&\vdots&\cdots&\vdots&\cdots&\vdots&\vdots\\

U(1)_{3,n-1}&\bold 0_a&\cdots&\bold 0_a&\bold 0_a&\cdots&\bold 0_a &\cdots&\bold 0_a
&\cdots&\bold 0_a&\bold 0_b&\cdots&\bold 1_b&\bold 0_b &\cdots&\bold 0_b&\cdots&\bold 1_b &\bold 0_b
\end{array}} \right) \ ,
\end{array}$$}
where $\bold 1_a$ and $\bold 0_a$ are row vectors of length $a$ with $1$ and $0$ in all directions, respectively, and similarly, $\bold 1_b$ and $\bold 0_b$ are row vectors of length $b$.
Note that the redundant row associated with $U(1)_{3,n}$ has been removed and the matrix $Q$ consists only of $l+n+m-1$ rows.\footnote{Topological invariants do not depend on the choice of the row removal.}

To evaluate the Euler number, we apply Eq.~\eqref{looped} with $m=c$,
\bea \label{formgen}\nn
\chi(M)&=&\frac{1}{|W|}\int_{\tilde X} {c(\cT \tilde X)}\wedge
\frac{e(\tilde \cN)}{c({\tilde \cN})}\wedge \frac{e({\bold\Delta})}{c({\bold\Delta})} \cr\cr \nn
&=&\frac{1}{l!\,m!\,n!} \left[\int_{\tilde X} \right]_{L_n=0}
 \underbrace{{\prod_{i=1}^{l}\prod_{j=1}^{m} (1-J_i +K_j)^a
 \wedge \prod_{j=1}^{m} \prod_{k=1}^{n} (1-K_j +L_k)^b}}_{c(\cT \tilde X)} \\
&&~~~~~~~~~~
\wedge\underbrace{\prod\limits_{i=1}^l \prod\limits_{k=1}^n
\left( \frac{-J_i+L_k }{1-J_i+L_k}\right)^c}_{e(\tilde {\mathcal N})
\,\wedge \, c(\tilde {\mathcal N})^{-1}} \\ \nn
&&~~~~~~~~~~
\wedge \underbrace{\frac{\prod\limits_{i \neq i'}^l (J_i - J_{i'})
\wedge\prod\limits_{j\neq j'}^m (K_j - K_{j'})
\wedge \prod\limits_{k\neq k'}^n (L_k - L_{k'})}{\prod\limits_{i\neq i'}^l (1+J_i - J_{i'})
 \wedge\prod\limits_{j\neq j'}^m (1+K_j-K_{j'}) \wedge\prod\limits_{k\neq k'}^{n}
 (1+L_k-L_{k'})}}_{e(\bold \Delta)\, \wedge\, c(\bold \Delta)^{-1}} \ ,
 \eea
where $J_{i=1, \cdots, l}$, $K_{j=1, \cdots, m}$
and $L_{k=1, \cdots, n-1}$ are the K\"ahler forms arising from the
three sets of $U(1)$'s, respectively.
Note that for symmetry of integrand, a formal variable $L_n$
has been introduced, which should, in the end, be taken to vanish, as indicated
by the integration symbol $\left[\int_{\tilde X} \right]_{L_n=0}$.
Then, inserting the intersection numbers turns the integral~\eqref{formgen}
over the manifold $\tilde X$ to an equivalent contour integral around the origin,
just as in Eq.~\eqref{contour}.

Similarly, the refined Euler character can be evaluated by applying
Eq.~\eqref{looped} with $m={\rm Td}\wedge {\rm ch}_\xi^*$,
\bea\label{formgen2} \nn
\chi_\xi (M)&=& \frac{1}{|W|} \int_{\tilde X}{{\rm Td}(\cT \tilde X)}
\wedge {{\rm ch}_\xi (\cT^* \tilde X)}\wedge
\frac{e(\tilde \cN)}{{\rm Td}({\tilde \cN})\wedge{\rm ch}_\xi({\tilde \cN^*})}
\wedge \frac{e({\bold\Delta})}{{{\rm Td}({\bold\Delta})}\wedge{{\rm ch}_\xi({\bold\Delta^*})} } \cr\cr
&=& \frac{1}{l!\,m!\,n!} \left[\int_{\tilde X}\right]_{L_n=0}\;
\prod\limits_{i=1}^l \prod\limits_{j=1}^m \left(\frac{-J_i+K_j}{1-e^{J_i-K_j}}\right)^a
\prod\limits_{j=1}^m \prod\limits_{k=1}^n \left(\frac{-K_j+L_k}{1-e^{K_j-L_k}}\right)^b \cr
&&~~~~~~~~~~~~\wedge\frac{1}{(1+\xi)^{l+m+n-1}}
\, {\prod_{i=1}^l \prod_{j=1}^m (1+\xi e^{-J_i + K_j})^a\prod_{j=1}^m \prod_{k=1}^n (1+\xi e^{-K_j + L_k})^b}  \\
&&~~~~~~~~~~~~\wedge\prod\limits_{i=1}^l \prod\limits_{k=1}^n
\left( \frac{1-e^{J_i-L_k}}{1+\xi e^{J_i-L_k}} \right)^c \cr
&&~~~~~~~~~~~~\wedge\prod\limits_{i\neq i'}^l \frac{1-e^{J_i-J_{i'}}}{1+\xi e^{J_i - J_{i'}} }
\prod\limits_{j\neq j'}^m \frac{1-e^{K_j-K_{j'}}}{1+\xi e^{K_j - K_{j'}}}\prod\limits_{k\neq k'}^n
\frac{1-e^{L_k-L_{k'}}}{1+\xi e^{L_k - L_{k'}}} \ ,
\eea
where the integration symbol $\left[\int_{\tilde X} \right]_{L_n=0}$ means that the
formal variable $L_n$ in the integrand is set to zero, as in the unrefined case~\eqref{formgen}.

For the rest of this section, we apply the index formulae~\eqref{formgen}
and~\eqref{formgen2} to two triangular, non-Abelian examples that illustrate
mutation equivalence and non-trivial quiver invariant, respectively.

\subsubsection{Consistency Check: Quiver Mutation}
\begin {figure}[h]
\centering
\includegraphics[scale=0.4]{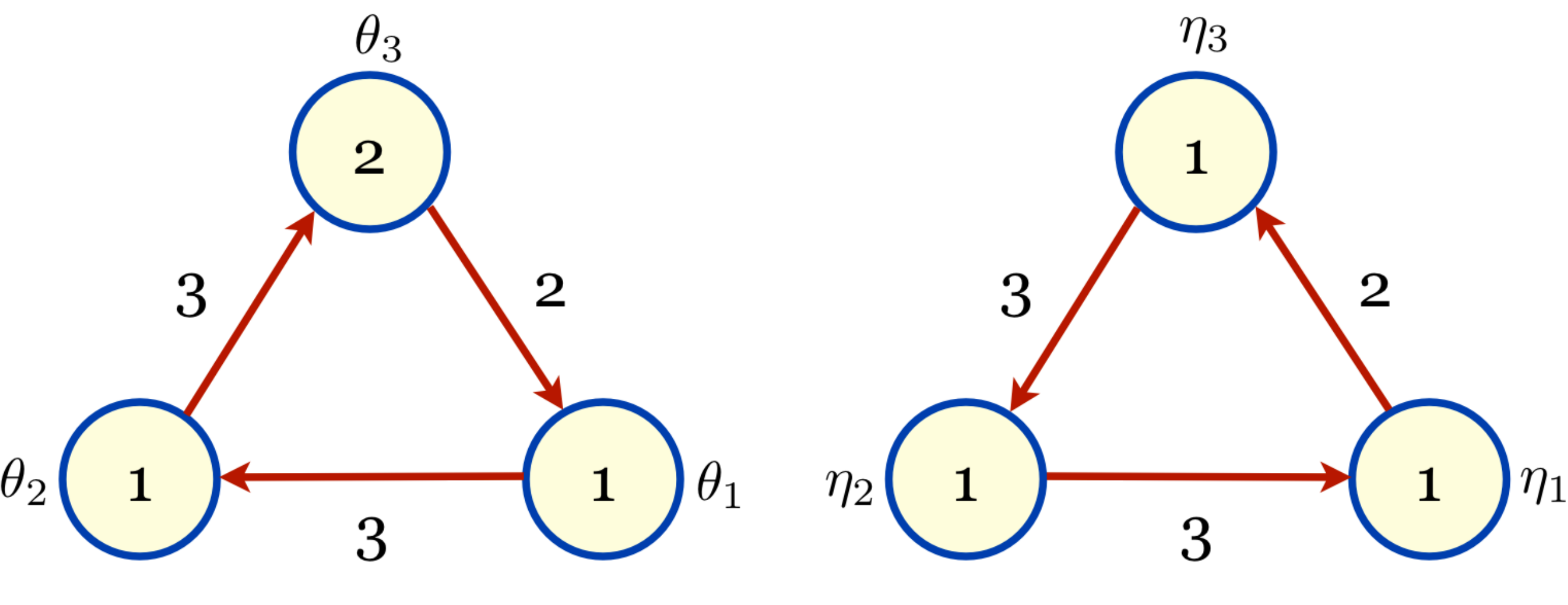}
\caption{ A non-Abelian triangular quiver (left) and an Abelian quiver (right);
the former is obtained by mutating node 3 of the latter.}\label{f:ex-mutation}
\end{figure}

Let us consider the non-Abelian quiver in Fig.~\ref{f:ex-mutation} (left),
in the branch where $\theta_1 >0$ and $\theta_2 <0$ so that the three arrows from node 1 to node 2 vanish.
The topology of the corresponding moduli space also depends on the sign of
$\theta_3$ and we consider the case where $\theta_3>0$. 
The Euler number and the refined Euler character are in turn obtained by
applying Eqs.~\eqref{formgen} and~\eqref{formgen2}, 
\bea \label{mutation1}
\chi(M)&=& 6 \ , \\ \label{mutation2}
\chi_\xi (M) &=& 1-4\xi+\xi^2\ .
\eea

On the other hand, the non-Abelian quiver in Fig.~\ref{f:ex-mutation}
on the left is mutation equivalent to the Abelian quiver on the right; the
former in the given branch arises from mutating node 3 of the latter
in the branch where $\eta_1 >0$ and $\eta_3<0$ so that  the two arrows from node 1 to node 3 vanish.

One can easily confirm that these are the correct branches by
transforming the FI constants under the mutation,
\bea
\theta_1&=&\eta_1 \ , \cr
\theta_2 &=& \eta_2 + 3 \eta_3 \ , \cr
\theta_3&=&-\eta_3 \ .
\eea
The Abelian index computations turn out to give exactly the same
results as in Eqs.~\eqref{mutation1} and~\eqref{mutation2}; this
provides a consistency check for the non-Abelian indices as the
two moduli spaces are mutation equivalent.

\subsubsection{Quiver Invariants Revisited}
\begin {figure}[h]
\centering
\includegraphics[scale=0.3]{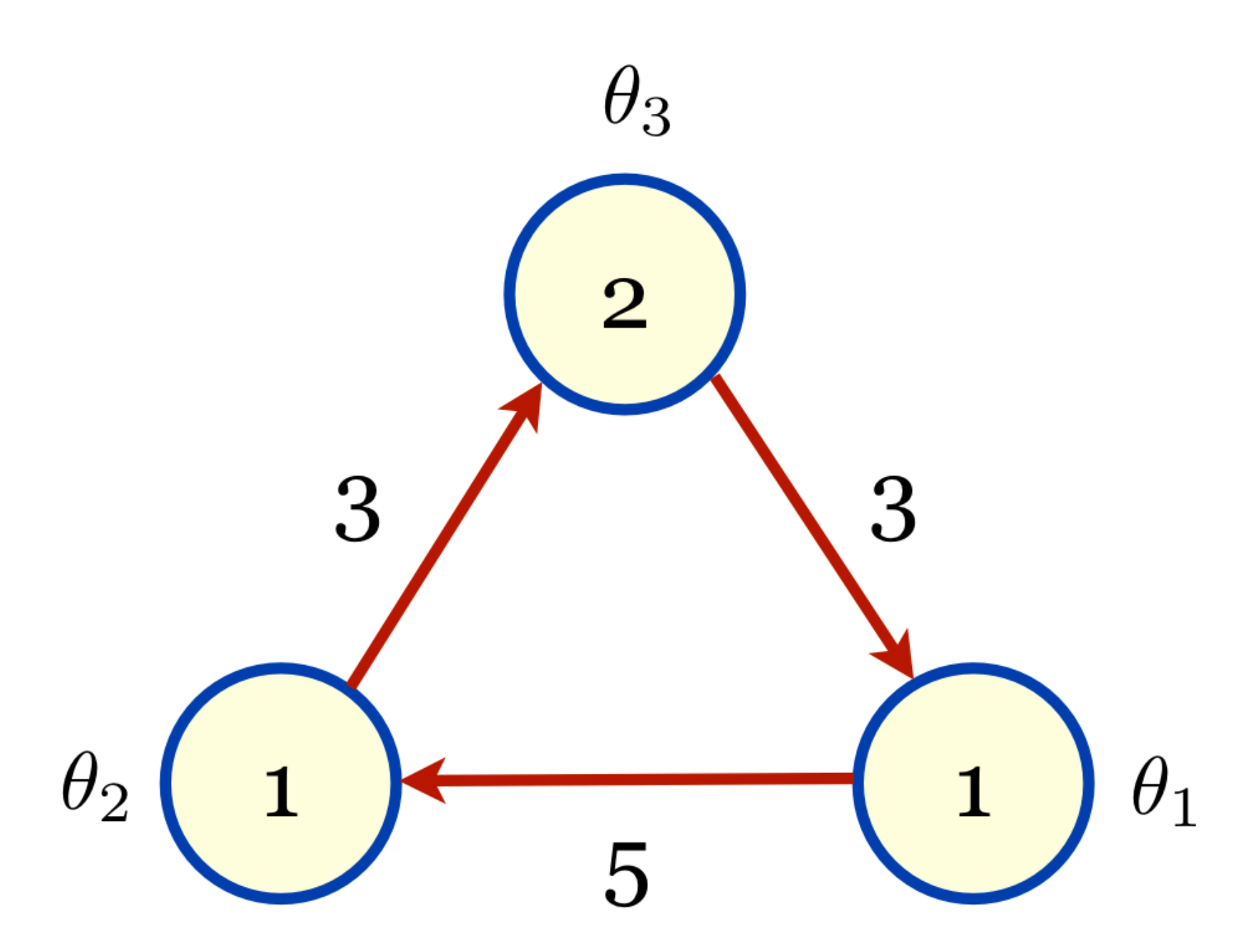}
\caption{A simplest non-Abelian quiver with intrinsic Higgs states }\label{f:ex-intrinsic}
\end{figure}

Let us take the triangular quiver in Fig.~\ref{f:ex-intrinsic},
which, due to the symmetry, has essentially two different branches:
(a) $\theta_1<0, \theta_3>0$ where the three arrows from node 3 to
node 1 vanish, and (b) $\theta_1>0, \theta_2<0$ where the five arrows
from node 1 to node 2 vanish.
The Euler number and the refined Euler character can be evaluated by
applying Eqs.~\eqref{formgen} and~\eqref{formgen2},
\beq
\begin{array}{rcl}
\chi(M_a)&=& 6 \  \\
\chi_\xi (M_a)&=& 6  \ ,
\end{array}
\text{~~~and~~}
\begin{array}{rcl}
\chi(M_b)&=& 9 \  \\
\chi_\xi (M_b)&=& 1-7\xi+\xi^2\ ,
\end{array}
\eeq
in the branches (a) and (b), respectively. 
Consequently, we have the following refined Higgs phase indices
\beq \label{higgs:ex-intrinsic}
\Omega_{\rm Higgs}^{(a)}(y)=6 \text{~~~and~~} \Omega_{\rm Higgs}^{(b)}(y)= \frac{1}{y^2} +7 + y^2 \ .
\eeq

The equivariant Coulomb phase indices, on the other hand, can be
separately computed, along the lines of Ref.~\cite{Manschot:2012rx,Manschot:2013sya}, as
\beq \label{coulomb:ex-intrinsic}
\Omega_{\rm Coulomb}^{(a)}(y)= 1+\Omega_{\rm Intrinsic}^{(a)}
\text{~~~and~~}
\Omega_{\rm Coulomb}^{(b)}(y)= \frac{1}{y^2}+2+y^2 + \Omega_{\rm Intrinsic}^{(b)} \ ,
\eeq
where $\Omega_{\rm Intrinsic}$ encodes the possibility of intrinsically
Higgs states that cannot be directly counted via the Coulomb approach.
By comparing Eqs.~\eqref{higgs:ex-intrinsic} and~\eqref{coulomb:ex-intrinsic},
we see
\beq
\Omega_{\rm Intrinsic}^{(a)}=\Omega_{\rm Intrinsic}^{(b)} = 5 =\Omega_{\rm Invariant}\ ,
\eeq
and find that the notion of the intrinsic Higgs states as chamber-independent invariant of
a quiver, first observed in Abelian quivers~\cite{Lee:2012sc,Lee:2012naa,Bena:2012hf},
manifests again in this non-Abelian example.\footnote{$\Omega_{\rm Invariant} = \Omega_S$
in the notation of Refs.~\cite{Manschot:2012rx,Manschot:2013sya}, where these were left as an unknown input data.}

\section{Abelianized Higgs Index is a Partition Sum}

\subsection{Coulomb Index as a Partition Sum}

With $\sum_A\gamma_A$ collection of BPS particles, with the intrinsic degeneracies
$\Omega(\gamma_A)$'s, the Coulomb index associated with the multi-particle BPS wavefunction
can be computed as
\begin{eqnarray}\label{C}
\Omega^-\left(\sum \gamma_A\right)&=& \Omega^+\left(\sum \gamma_A\right)\cr\cr
 &+& (-1)^{\sum_{A>B}\langle\gamma_A,\gamma_B\rangle+n-1}
\frac{\prod_A  \bar\Omega(\gamma_A)}{|\Gamma|}\int_{\cal M} ch({\cal F})\wedge {\cal A}({\cal M})\cr\cr
&+& (-1)^{\sum_{A'>B'}\langle\gamma_{A'}',\gamma_{B'}'\rangle+n'-1}
\frac{\prod_{A'} \bar \Omega(\gamma_{A'}')}{|\Gamma'|}\int_{{\cal M}'}
ch({\cal F'})\wedge {\cal A}({\cal M}')\cr\cr
&+&(-1)^{\sum_{A''>B''}\langle\gamma_{A''}'',\gamma_{B''}''\rangle+n''-1}
\frac{\prod_{A''} \bar \Omega(\gamma_{A''}'')}{|\Gamma''|}\int_{{\cal M}''}
ch({\cal F''})\wedge {\cal A}({\cal M}'')\cr\cr
&+&\cdots
\end{eqnarray}
where the charge $\sum_A\gamma_A$ is assumed to be primitive. $\Omega^\pm$
denote the indices on the two sides of marginal stability wall. When we
cross the marginal stability wall, the Coulomb vacuum manifold ${\cal M}$
and its submanifolds ${\cal M}', {\cal M}'',$ etc. become noncompact, and all
the quantum ground states become non-normalizable. Only the first term $\Omega^+(\sum_A\gamma_A)$ may be present. This is how wall-crossing occurs
from the Coulomb viewpoint.

The sum is over partitions of the total charge,
$$\sum_{A=1}^n \gamma_A=\sum_{A'=1}^{n'}\gamma'_{A'}
=\sum_{A''=1}^{n''}\gamma''_{A''}=\cdots$$
such that $\gamma'_{A'}$ etc. are generally non-negative-integer linear
combination of $\gamma_A$'s. Note that $n>n'$, meaning that in each of the
partition, we have fewer number of particles than in the original problem.
The barred $\Omega$'s are the so-called rational
invariant
$$\bar\Omega(\gamma)=\sum_{p\vert\gamma} \Omega(\gamma/p)/p^2$$
defined as a sum over all divisors of the charge in question.
When  $\sum_A\gamma_A$ is not primitive, it suffices to replace
$\Omega^\pm\left(\sum \gamma_A\right)$ by its rational counterpart.
Finally $\Gamma'$ is a set of permutation groups that mixes up identical
charges among $\gamma_A'$'s.
For further details of the formula, such as the nature of space ${\cal M}$'s
and the magnetic field strengths ${\cal F}$'s, we direct readers to Ref.~\cite{Kim:2011sc}.

This wall-crossing formula actually incorporates possibility of BPS states
of the same charge arising from different constituent particles, i.e.,
from different quivers. For a single quiver, such a sum is due to the
Weyl projection, or equivalently from the quantum statistics of
indistinguishable particles. Starting with a particular quiver
with a rank-$r_v$ node of a primitive charge $\gamma_v$,  a partition,
$r_v=\sum_{a_v=1}^{l_v} r_{v,a_v}$, contributes a term proportional to
$$\frac{1}{|\Gamma(\{r_{v,a_v}\})|}\prod_{a_v=1}^{l_v} \Omega(\gamma_v)/r_{v,a_v}^2$$
where  $\Gamma(\{r_{v,a_v}\})$ is subgroup of the permutation group $S(r_v)$,
or the Weyl group, that survives when some of $r_{v,a_v}$'s equal. The
effective moduli space ${\cal M}'$ for such a partition is that of
$l_v$ particles of charges $\gamma'_{A'}=r_{v,a_v} \, \gamma_v$ for $a_v =1, \cdots,  l_v$, etc.,
which in turn is a submanifold of ${\cal M}$.
The above general wall-crossing formula can be rebuilt from this
by allowing different quivers contributing to the same
charge states \cite{Kim:2011sc}.\footnote{
Ref.~\cite{Manschot:2010qz} also arrived at the same formula, again utilizing quantum statistics
but in a rather different manner. In the approach of Ref.~\cite{Manschot:2010qz},
one factor $1/p$ in the rational invariant arises from statistics while
another  $1/p$   arises from the assumption that basic building blocks of the index
are linear with respect to  the pairwise Schwinger products. The latter
assumption actually fails for quivers with scaling regime, just as the approach
of Ref.~\cite{Kim:2011sc} cannot be trusted either in the presence of the scaling regime.}

Recall that, at least for  cases {\it without} scaling regimes, the
Higgs and the Coulomb answers  have been shown
to be equivalent~\cite{Denef:2002ru, Sen:2011aa}. This implies that
Higgs index of some non-Abelian quivers can also be decomposed into a sum
of indices of finitely many Abelian quivers, which are obtained by partitions
of each non-Abelian nodes, say, of rank $r_v$, as $r_v=\sum_{a_v} r_{v,a_v}$.  While
quite natural and proven rigorously on the Coulomb side,
the physical or mathematical origin of such a decomposition is
quite opaque in the Higgs side.
For the quivers with a loop and the scaling regime, furthermore,
the equivalence between  Coulomb and Higgs sides no longer holds in
general. Higgs side is more comprehensive and one finds states
missed by the Coulomb ``phase" computation.
So one might wonder whether the partition sum representation
is still possible for such quivers at all.

We wish to argue
that our Abelianization routine for computing Higgs ``phase" index
naturally leads again to another sum over partitions of the total charge,
which can be viewed as the Higgs counterpart of (\ref{C}).
In the next subsection, we shall study our Abelianization procedure and, for some simple cases without intrinsic Higgs states, compare the resulting partition sum with those found in Coulomb ``phase" computation.
The comparison will be made term by term and a complete agreement found for the examples.

\subsection{Is the Higgs Index Also a Partition Sum?}

Let us revisit the general Abelianization prescription for computing
the refined Higgs index. In sections 2 and 3, we proposed and tested
for some examples that it can be computed starting with the Cartan
data of the quiver as
\beq\label{chiref22}
\Omega(y)=
\frac{1}{|W|}\int_{\tilde X} {\omega_y(\cT \tilde X)}\wedge
\frac{e(\tilde \cN)}{\omega_y({\tilde \cN})}\wedge \frac{e({\bold\Delta})}{\omega_y({\bold\Delta})}\ ,
\eeq
where $\tilde M$ and $\tilde X$ are the
would-be Higgs moduli space and the ambient D-term variety, respectively,
of the Abelianized quiver, and $\tilde{\cal N}$ is
the normal bundle of $\tilde M$ embedded in $\tilde X$. The gauge group
of the Abelianized quiver is the Cartan subgroup $T$ of
the non-Abelian quiver but the chiral multiplet contents are kept intact.
 Note that for the Abelianized quiver, the refined index is  given
by integration of the first two factors in the
integrand~\eqref{chiref22} without the last factor depending on $\bold \Delta$.

Recall that the vector bundle $\bold \Delta={\bold \Delta}^*$ is the sum of line bundles,
associated with ``off-diagonal" part of the gauge group $G$,
$$\bigoplus_{\alpha\in \Delta} {\cal L}_\alpha = \left(\bigoplus_{\alpha\in \Delta^+}
{\cal L}_\alpha\right)\bigoplus\left( \bigoplus_{\alpha\in \Delta^-} {\cal L}_\alpha\right) $$
with fiber  $G/T$, which gives
\beq\label{vdm}
\frac{ e({\bold \Delta})   }{\omega_y(\bold \Delta) }=
\prod_{\alpha \in \Delta} \frac{1-e^{-c_1({\cal L}_\alpha)}}{y e^{ -c_1({\cal L}_\alpha)}-y^{-1}}=
\prod_{\alpha \in \Delta^+}\left(1-\delta_\alpha\right) \ ,
\eeq
which defines $\delta_\alpha$ for each positive root $\alpha$ as
\beq\label{delta}
\delta_\alpha \equiv \frac{(y - y^{-1})^2}{y^2+y^{-2} - 2\cosh(c_1(\cL_\alpha))} \ .
\eeq
Note that the expression~\eqref{vdm}, when expanded, has $2^{|\Delta^+|}$ terms of the form
\beq\label{deltaExpand}
\delta(\mathcal I)\equiv (-1)^{|\mathcal I|}
\prod_{\alpha \in \mathcal I} \delta_{\alpha} \ ,
\eeq
where $\mathcal I \subset \Delta^+$ is a collection of positive roots.

It is clear that the very first term in the expansion of (\ref{vdm}),
i.e., the term without any $\delta_\alpha$ factors, computes in (\ref{chiref22}) the Higgs
index of the Abelianized quiver $\tilde \cQ$, divided by $|W|$.
Recall that we started with a quiver $\cQ$ with the gauge group $\prod_{v=1}^N U(r_v)$,
where $v$ labels the $N$ nodes, and then obtained an Abelian quiver $\tilde \cQ$
of rank  $r+1=\sum_v r_v$ by replacing $\prod_v U(r_v)$ in $\cQ$ by
$\prod_v U(1)^{r_v}$ and maintaining the chiral field contents intact.
$\tilde M$ is precisely the Higgs moduli space of $\tilde \cQ$
in the chamber determined by $\theta$'s inherited from $\cQ$.

A given positive root $\alpha$ of $\prod_v U(r_v)$ connects a distinct and
unique pair of Cartan generators
in $\prod_v U(1)^{r_v}$, which tells us that for any given subset
$\mathcal I \subset \Delta^+$ we can associated an unordered partition
${\cal P}_{\cal I}$.
One merely counts Cartan generators, connected pairwise by elements of ${\cal I}$,
and call the resulting integers, $ r_{v,a_v}$.
Two extreme examples are the maximal partition,
\beq
{\mathcal P}_{\slashed o}=(\{\underbrace{1, 1, \cdots, 1}_{r_1~\text{copies of $1$}}\};
\{\underbrace{1, 1, \cdots, 1}_{r_2~\text{copies of $1$}}\}\cdots;
\{\underbrace{1, 1, \cdots, 1}_{r_N~\text{copies of $1$}}\}) \ ,
\eeq
corresponding to the empty subset ${\cal I}=\slashed o$, and the minimal,
or trivial, partition
\beq
 \mathcal P_{\Delta^+} = (\{r_1\}, \{r_2\}, \cdots, \{r_N\}) \ .
\eeq
for ${\cal I}=\Delta^+$.
Note that there are
in general many different ${\cal I}$'s
that map to the same unordered partition ${\cal P}$.

For each (unordered) partition ${\cal P}$, we may associate an Abelian quiver
$\cQ_{\cal P}$. The Abelian quiver $\tilde \cQ$ we encountered many time
already is an example of this,
$$\tilde \cQ=\cQ_{\cal P_{\slashed o}} \ . $$
One way to picture $\cQ_{\cal P}$ for a given $\cP=(\{ r_{v,a_v}\})$ is to start with
$\tilde \cQ =\cQ_{\cal P_{\slashed o}} $, gather $U(1)$ nodes of
$\tilde \cQ$ according to the numbers $\{r_{v,a_v}\}$, and fuse
each such collection to a single Abelian node, of which FI constant is given by summing those of the fused nodes. In terms of the Coulomb side
picture, such a node corresponds to a single  center of (non-primitive) charge
$r_{v,a_v}\,\gamma_v$. Naturally $\cQ_{\cal P}$ has the gauge  group
$\prod_{v}\prod_{a_v} U(1)$,
typically of smaller rank than $r+1=\sum_v r_v=\sum_v \sum_{a_v} r_{v,a_v}$.
We keep the bi-fundamental chiral field contents intact, which is
accomplished as the intersection numbers are multiplied by a pair of $r_{v,a_v}$'s in an obvious manner:
$$
r_{v,a_v}\times  \langle \gamma_v, \gamma_{w} \rangle \times r_{w,b_{w}} \ .
$$

\begin {figure}[h]
\centering
\includegraphics[scale=0.38]{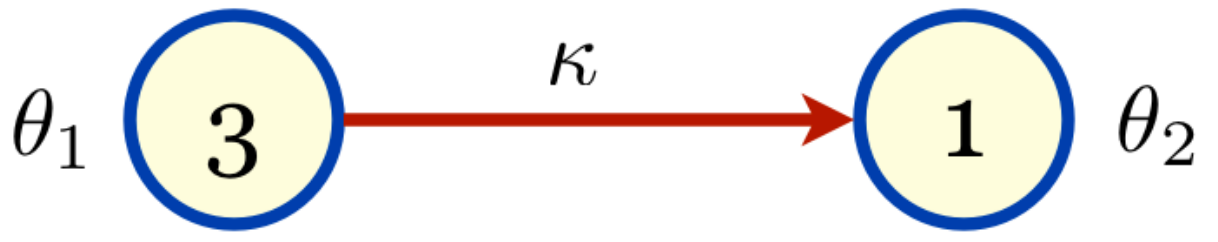}
\caption{Grassmannian quiver with $\bold d=(3,1)$ and linking number $\kappa$}
\label{f:ex-Gr-partition}
\end{figure}

\begin {figure}[h]
\centering
\includegraphics[scale=0.38]{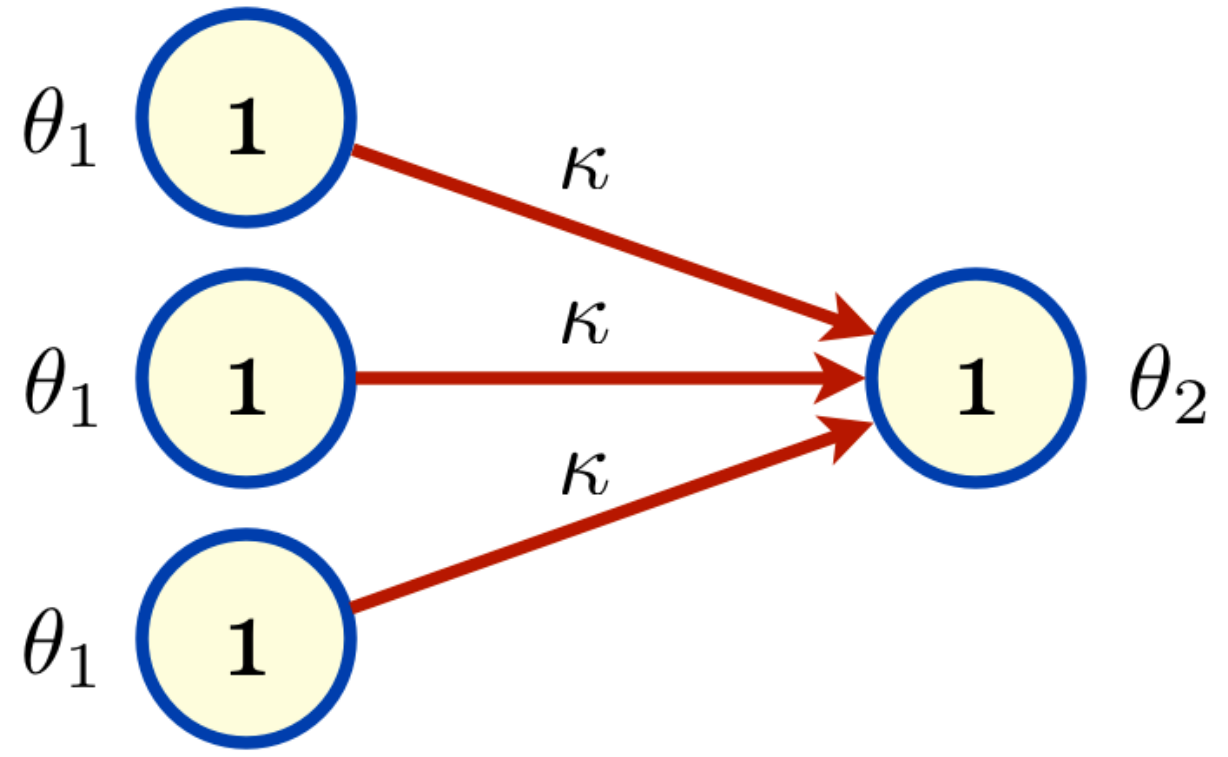}~~~
\includegraphics[scale=0.38]{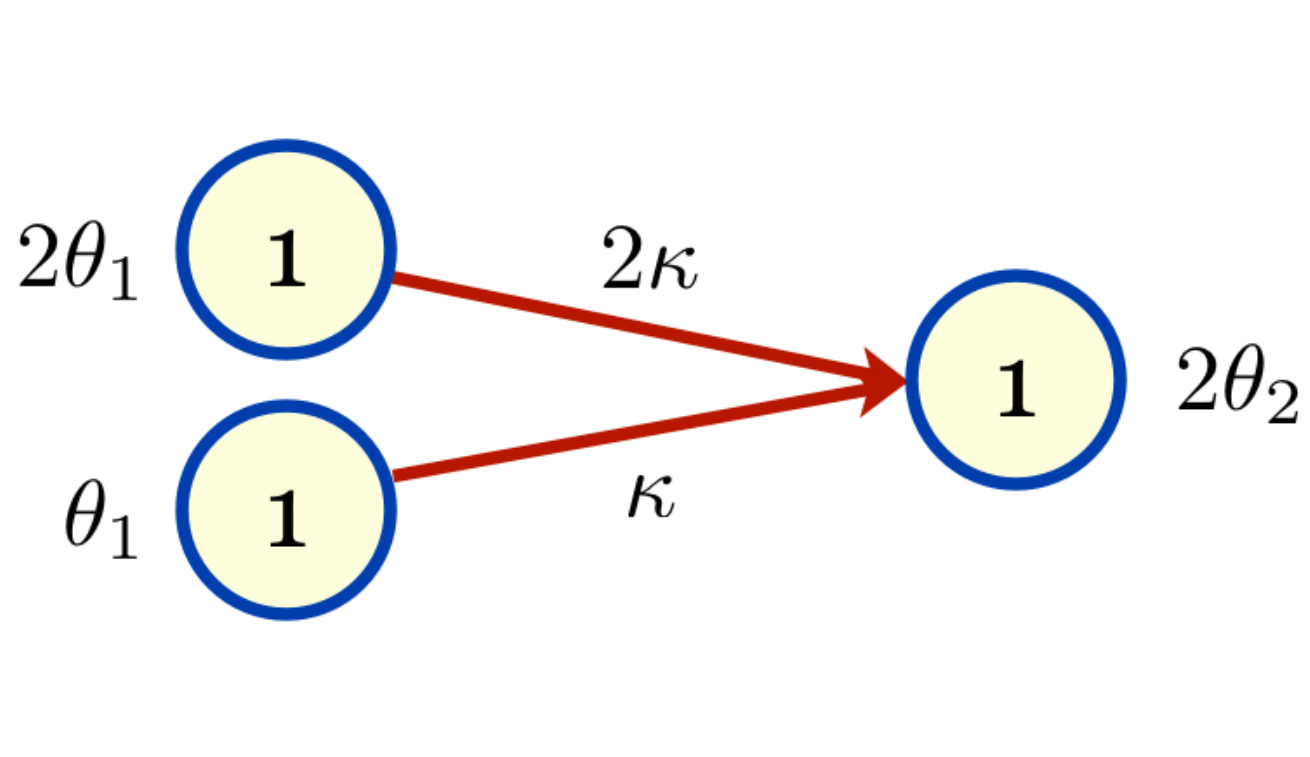}~~~
\includegraphics[scale=0.38]{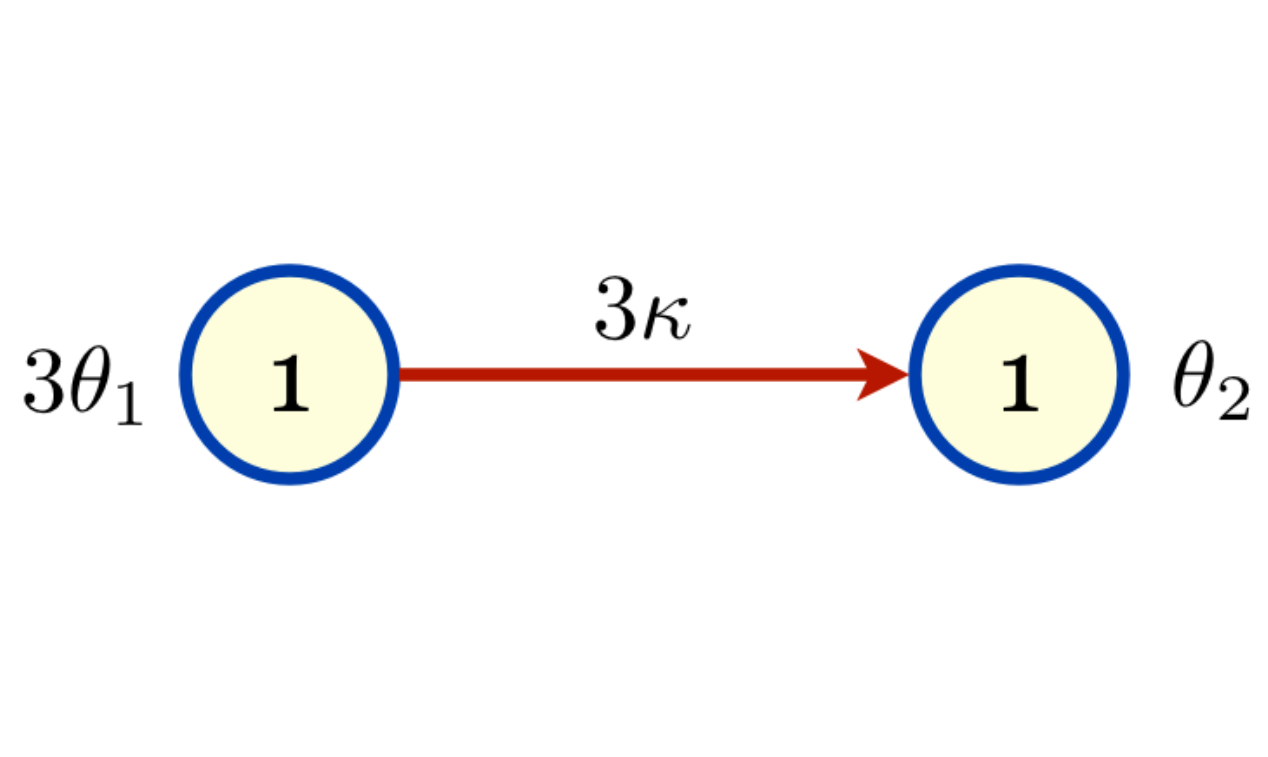}
\caption{The three Abelian quivers arising from the non-Abelian quiver in Fig.~\ref{f:ex-Gr-partition}; they correspond, respectively(from left to right), to the three partitions $\cP_i$ for $i=1,2,3$ given in Eq.~\eqref{Gr-partition}.  }
\label{f:ex-Gr-partition-Abelianized}
\end{figure}

As an illustration of the mapping $\cP \mapsto \cQ_\cP$, let us consider the Grassmannian quiver in Fig.~\ref{f:ex-Gr-partition}, with gauge group $G=U(3)\times U(1)$, that has the dimension vector $\bold d=(3,1)$.
The dimension vector admits the following three partitions,
\bea \nn \label{Gr-partition}
\cP_1&=& (\{1,1,1\}; \{1\}) \ , \\
\cP_2&=& (\{1,2\}; \{1\}) \ , \\ \nn
\cP_3&=& (\{3\};\{1\}) \ ,
\eea
which, according to the rule explained in the previous paragraph, correspond, respectively, to the three quivers depicted in Fig.~\ref{f:ex-Gr-partition-Abelianized}.

To state the conjecture, we now come back to the contribution from $\mathbf{\Delta}$ to
the refined Higgs index. Reorganizing the terms  as
\beq
\frac{ e({\bold \Delta})   }{\omega_y(\bold \Delta) }=
\prod_{\alpha \in \Delta^+}\left(1-\delta_\alpha\right)
=
\sum_{\cal P}\sum\limits_{\mathcal P_{\mathcal I}
={\mathcal P},\,\mathcal I \subset \Delta^+} \delta(\mathcal I)\ ,
\eeq
we have a sum over partitions for the refined index as
\beq\label{chiref3}
\Omega(y)[\cQ]=\sum_{\cal P} \left(
\frac{1}{|W|}\int_{\tilde X} \left[{\omega_y(\cT \tilde X)}\wedge
\frac{e(\tilde \cN)}{\omega_y({\tilde \cN})}\wedge \sum\limits_{\mathcal P_{\mathcal I}
={\mathcal P},\,\mathcal I \subset \Delta^+} \delta(\mathcal I)\right]\right)\ .
\eeq
For a general quiver, we claim that each and every term in the sum over ${\cal P}$
represents contribution from the Abelian quiver $\cQ_{\cal P}$ defined above.

As we noted earlier, a partition sum of similar kind has been rigorously
demonstrated on the Coulomb side computation, which is reliable for
quivers without loops. As the Higgs index and the Coulomb index equal in these cases,
a partition sum does already exist in the Higgs side as well. What we
claim is that, for such general quivers, our partition sum coincides
exactly and term--by-term with this physically motivated partition sum.
Borrowing from these works, then, our conjecture can be stated as
\beq\label{pquiver}
\frac{1}{|W|}\int_{\tilde X} \left[{\omega_y(\cT \tilde X)}\wedge
\frac{e(\tilde \cN)}{\omega_y({\tilde \cN})}\wedge \sum\limits_{\mathcal P_{\mathcal I}
={\mathcal P},\,\mathcal I \subset \Delta^+} \delta(\mathcal I)\right]
\;=\;\;c({\mathcal P};y)\times \Omega(y)[Q_{\cal P}]
\eeq
for each unordered partition $\cP$, where
\beq\label{prefactor}
c(\mathcal P;y) \equiv  \frac{1}{|\Gamma(\mathcal P)|}\prod\limits_{v=1}^N
\prod\limits_{a_v=1}^{l_v}\frac{1}{r_{v, a_v} } \frac{y-y^{-1}}{y^{r_{v, a_v}} - y^{-r_{v, a_v}}} \ ,
\eeq
is a well-established universal factor that appears in the Coulomb phase
wall-crossing formula. For the nonequivariant limit, terms in the product
here reduce to $\pm 1/r_{v,a_v}^2$ we already encountered at the
top of this section. Although $c(\mathcal P;y) $ was found in the study of
quivers without loops, its origin lies entirely in the quantum statistics or equivalently
the Weyl groups and the same formula should be applicable to general quivers.

Again, to illustrate our proposal for the new partition-sum structure, let us revisit the Grassmannian example in Fig.~\ref{f:ex-Gr-partition}. Firstly, the gauge group $G=U(3)\times U(1)$ has three positive roots, which we may denote as $\Delta^+ = \{\alpha_{12}, \alpha_{13}, \alpha_{23}\}$.
Then, for instance, the subset $\cI = \{\alpha_{12}\} \subset \Delta^+$ maps to the partition $\cP_\cI = (\{1,2\}; \{1\})$ and hence, to the second quiver in Fig.~\ref{f:ex-Gr-partition-Abelianized}, as the two out of the three nodes arising from the rank-$3$ node of the original non-Abelian quiver, are fused together via the single element $\alpha_{12}$ in $\cI$. It is easy to see that the eight subsets of $\Delta^+$ can be grouped into the following three,
\bea \nn\label{I-regroup}
\cI_1 &=& \slashed o \ , \\
\cI_2 &=& \{\alpha_{12}\}, \{\alpha_{13}\}, \{\alpha_{23}\} \ , \\ \nn
\cI_3 &=& \{\alpha_{12}, \alpha_{13}\}, \{\alpha_{12}, \alpha_{23}\}, \{\alpha_{13}, \alpha_{23}\}, \{\alpha_{12}, \alpha_{13}, \alpha_{23}\} \ ,
\eea
so that $\cI_i$ map to $\cP_i$ for $i=1,2,3$ and, in turn, to the three quivers in Fig.~\ref{f:ex-Gr-partition-Abelianized}, respectively (from left to right). Our conjecture~\eqref{pquiver} asserts, for instance, that the left hand side, which is a sum over the three subsets $\cI_2$ in Eq.~\eqref{I-regroup}, should equal the Higgs phase index on the right hand side for the second quiver in Fig.~\ref{f:ex-Gr-partition-Abelianized}, multiplied by the appropriate prefactor~\eqref{prefactor}, which in this case is
\beq
c(\cP_{\cI_2}; y) = \frac{1}{2} \frac{1}{y+y^{-1}} \ .
\eeq

To summarize, we rewrote the Abelianization formula for refined Higgs index
in terms of a partition sum. Each summand, labeled by ${\cal P}$, is then
conjectured to compute refined Higgs index of a specific Abelian quiver
$Q_{\cal P}$, up to universal factor $c(\mathcal P;y)$. As noted many times,
 quivers without loops are known to admit an expansion via partition of the total
charge, and this was motivated, tested, and proven in the Coulomb description
and thus is trustworthy for such quivers. What our computation and
the conjecture suggests, as we will see below for the simplest classes of
non-Abelian quiver, is that (\ref{chiref3}) coincides with this existing
partition sum via  Eq.~(\ref{pquiver}), even though our partition-sum
expansion comes from an entirely mathematical manifestation in the Higgs description.

The real substance of this conjecture lies in that this phenomenon holds
for general quivers with loops as well. As the Abelianization procedure
works in the presence of F-terms, the partition sum (\ref{chiref3})
clearly holds as a mathematical statement, yet, whether Eq.~(\ref{pquiver})
holds is hardly clear, a priori. For general quivers with loop, another
conjectural form of such a partition sum has been proposed by Manschot, Pioline, and Sen
\cite{Manschot:2012rx,Manschot:2013sya}. However, this leaves
behind the counting of the so-called intrinsic Higgs states as undetermined input data.
Our formulae compute the Higgs index, including contributions from the quiver
invariants directly, and thus offer a self-complete routine for counting BPS states.

\paragraph{Two-Node Quivers: }

Let us prove the conjecture~\eqref{pquiver} for the Grassmannian quivers,
consisting of an Abelian and a rank-$r$ nodes connected by $\kappa$
bi-fundamental fields. See Fig.~\ref{f:ex-Gr-partition} for the case of $r=3$.
The Higgs moduli space is $Gr(r,\kappa)$.
The formula~\eqref{chiref22} is expanded as
\beq\label{chiGr}
\Omega(y)[Gr(r,\kappa)]= \frac{1}{|W|} \int_{\tilde X} \left[{w_y({\cal T}\tilde X )}\wedge
\left(\,\sum\limits_{\mathcal I \subset \Delta^+} \delta(\mathcal I)\right)\right]
\eeq
where $\delta(\mathcal I)$'s are as defined in Eqs.~\eqref{deltaExpand} and~\eqref{delta}:
\beq\nn
\delta(\mathcal I) = (-1)^{|\mathcal I|} \prod_{\alpha \in \mathcal I}  \delta_\alpha  \quad
\text{with} \quad \delta_\alpha = \frac{(y - y^{-1})^2}{y^2+y^{-2} - 2\cosh(c_1(\cL_\alpha))} \ .
\eeq
Note that the normal bundle pieces are missing as the quivers are tree-like.
More explicitly, by introducing the $r$ K\"ahler forms $J_{i=1, \cdots, r}$
of $\tilde X = (\IP^{\kappa-1})^r$, the expansion~\eqref{chiGr} can be rewritten as
\bea\label{chiGrExplicit} \nn
&&\Omega(y)[Gr(r,\kappa)] \cr\cr
&=&\frac{1}{|W|}\frac{1}{(y-y^{-1})^r} \int_{\tilde X} \left[\prod_{i=1}^r
\left(J_i\wedge \frac{ye^{-J_i}-y^{-1}}{1-e^{-J_i}} \right)^\kappa\right]
\wedge \left[\,\sum_{\mathcal I \subset \Delta^+} \delta(\mathcal I)\right] \\
&=& \frac{1}{|W|} \frac{1}{(y-y^{-1})^r} \oint \left[\prod_{i=1}^r {\rm d}J_i \right]
\left[ \prod_{i=1}^r \left(\frac{y e^{-J_i}-y^{-1}}{1-e^{-J_i}} \right)^\kappa\right]
\left[\,\sum_{\mathcal I \subset \Delta^+} \delta(\mathcal I)\right] \ ,
\eea
where the contour integrals are around the circles centered at origin and
the $(2 \pi i)^{-1}$ factor is implicit in each measure ${\rm d}J_i$. Our
conjecture states that terms in the final sum are associated,
via (\ref{pquiver}), with Abelian
quivers $\cQ_\cP$ labeled  by partition $\cP$ of $(r,1)$ ; for $r=3$, three
such Abelian quivers are found in  Fig.~\ref{f:ex-Gr-partition-Abelianized}.

A simplest way to verify the partition-sum structure~\eqref{pquiver} is to
consider the decomposition inductively on the rank $r$ of the non-Abelian node:
\begin{itemize}
\item For $r=1$ case, $\Delta^+$ is empty and thus the factor
$\sum_{\mathcal I \subset \Delta^+} \delta(\mathcal I)$ in the integrand of
Eq.~\eqref{chiGr} only has a single term, the unity, leading to the refined
Euler character of $X(=\tilde X)$ itself.

\item For $r=2$ case, $\Delta^+=\{\alpha_{12}\}$ is a singleton with
the unique positive root $\alpha_{12}$ of $U(2)$ and the factor
$\sum_{\mathcal I \subset \Delta^+} \delta(\mathcal I)$ in Eq.~\eqref{chiGr}
has two terms, $\delta(\slashed o)=1$ and $\delta(\Delta^+)=- \delta_{\alpha_{12}}$.
The former corresponds exactly to Eq.~\eqref{pquiver} for the maximal partition
$\mathcal P=(\{1,1\}; \{1\})$ with $c(\mathcal P;y)=\frac{1}{|W|}$.
One can go brute force and also verify explicitly that the latter
corresponds to Eq.~\eqref{pquiver} for the minimal partition $\cP=(\{2\}; \{1\})$.
There is a simpler way around to check this, however.
Note first that the both sides of relation~\eqref{pquiver}, when summed over
all partitions $\cP$, lead to one and the same invariant, the refined index of
$Gr(r,\kappa)$. As we have already checked the
validity of relation~\eqref{pquiver} for the maximal partition out of the two
available partitions, it should also hold for the remaining, minimal partition.

\item Let us suppose that the relation~\eqref{pquiver} holds for all
Grassmannian quivers with rank of the non-Abelian node less than $r$
and consider the $Gr(r, \kappa)$ quiver. Since the first factor
$$ \prod_{i=1}^r \left(\frac{y e^{-J_i}-y^{-1}}{1-e^{-J_i}} \right)^\kappa $$
of the integrand in Eq.~\eqref{chiGrExplicit} factorizes to
$r$ pieces, each depending on a single $J_i$ variable, the
inductive assumption can be used to show that the relation~\eqref{pquiver}
holds for all partitions except possibly for the minimal one, $\cP= (\{r\};\{1\})$.
Then we may apply the same argument illustrated above for $r=2$ case:
Since the both sides of~\eqref{pquiver} sum up to a common invariant and
the relation~\eqref{pquiver} is valid individually for all but the minimal
partition, it should also hold for the minimal partition.
\end{itemize}

\paragraph{A Simple Non-Abelian Cyclic Quiver:}

\begin {figure}[h]
\centering
\includegraphics[scale=0.4]{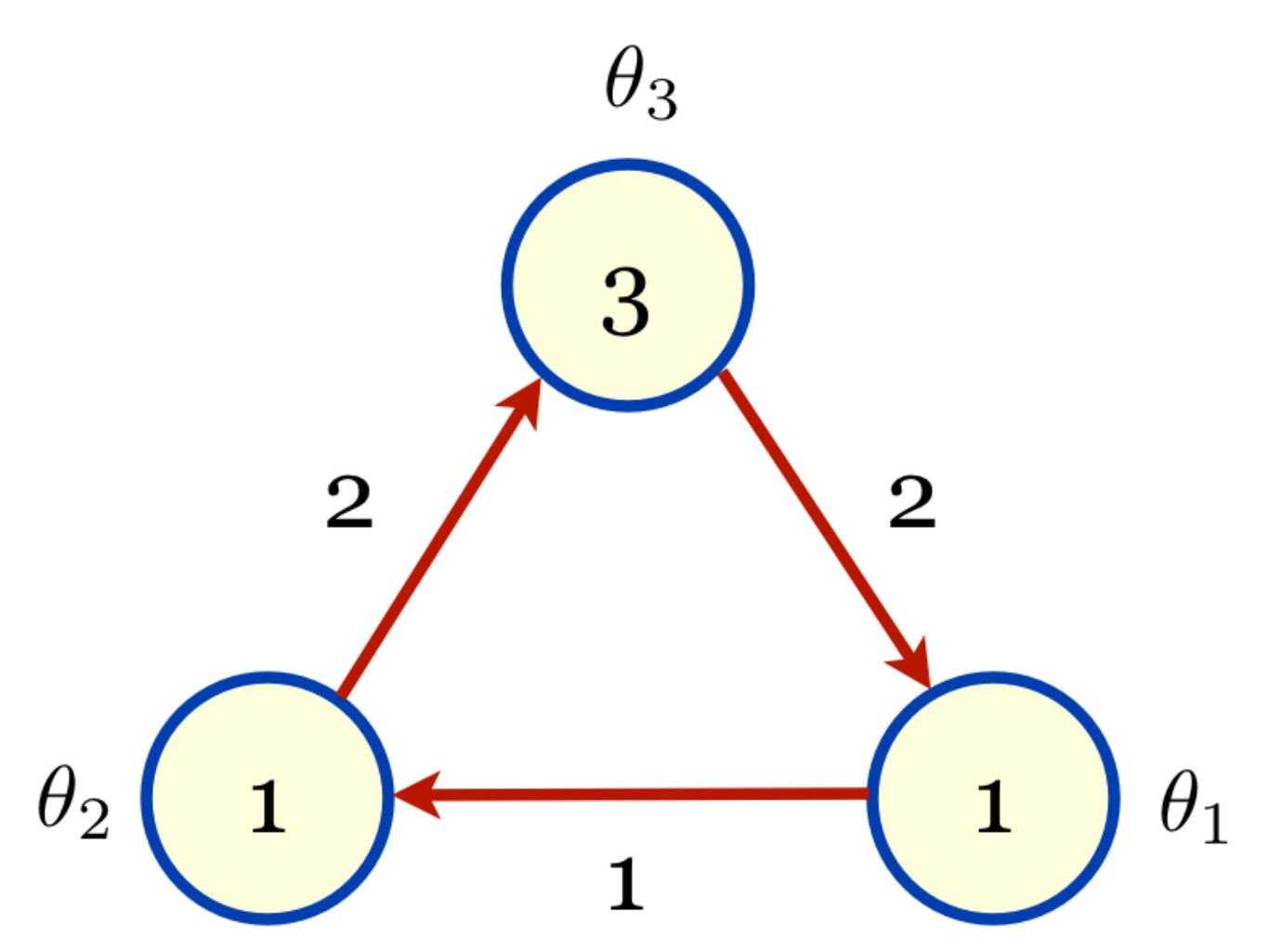}
\caption{A non-Abelian quiver with a rank-3 node and two Abelian nodes}\label{f:NonAbPartitionEx}
\end{figure}

As a next example, we take a non-Abelian cyclic and triangular
quiver in Fig.~\ref{f:NonAbPartitionEx}. For the simplest such
case, with rank 2 instead of rank 3 node at the top, the conjectured
identity \eqref{pquiver} holds somewhat trivially, once
that it holds for $\cP_{\slashed o}$ term, because there is only
two possible partitions. Rank 3 case therefore offers the first
nontrivial test. Also this case is a significant departure from
the Grassmannian example above, because the quiver
in question comes with a loop, and therefore the quiver dynamics
has a superpotential.

\begin {figure}[h]
\centering
\includegraphics[scale=0.35]{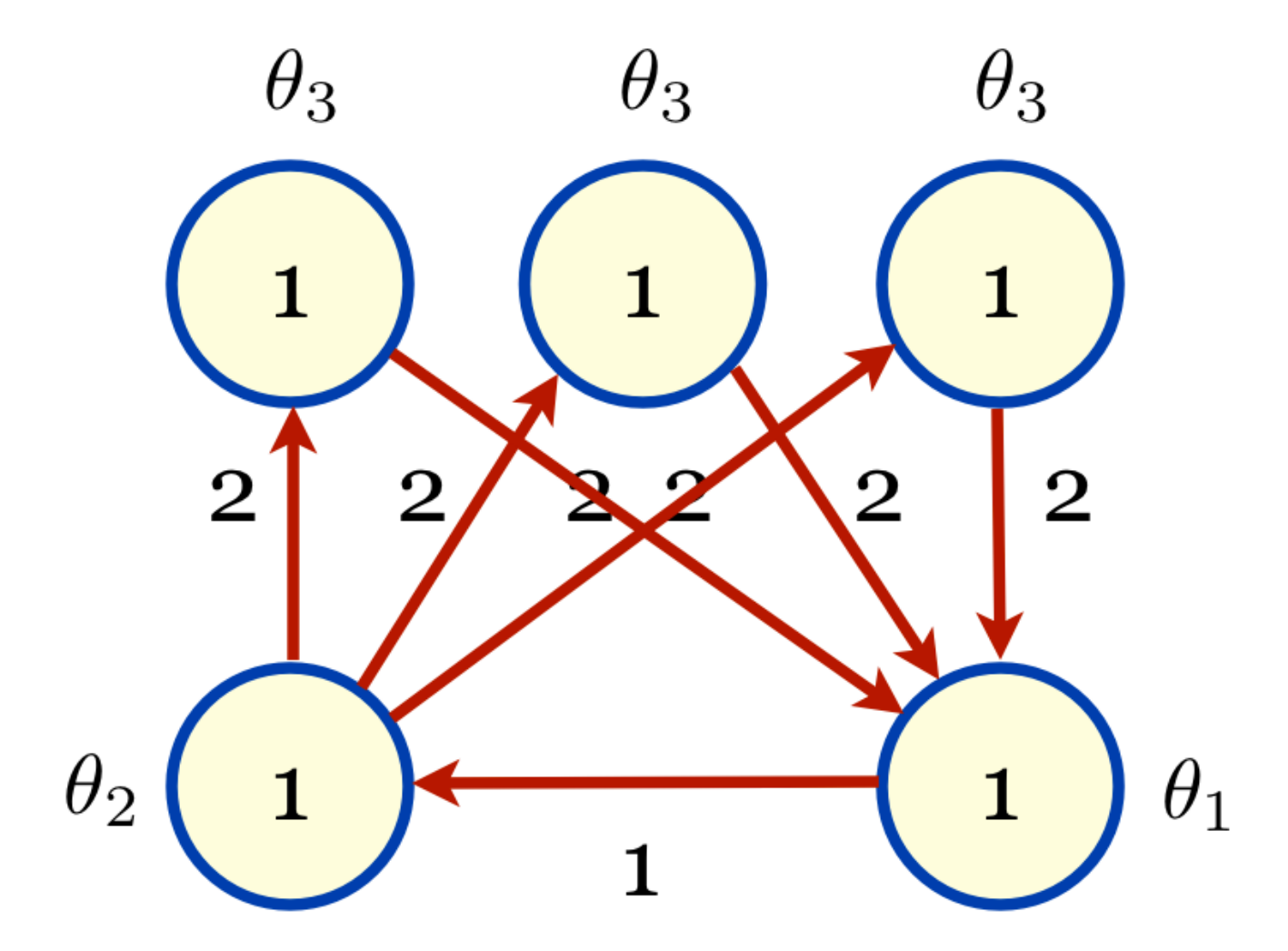}
\includegraphics[scale=0.35]{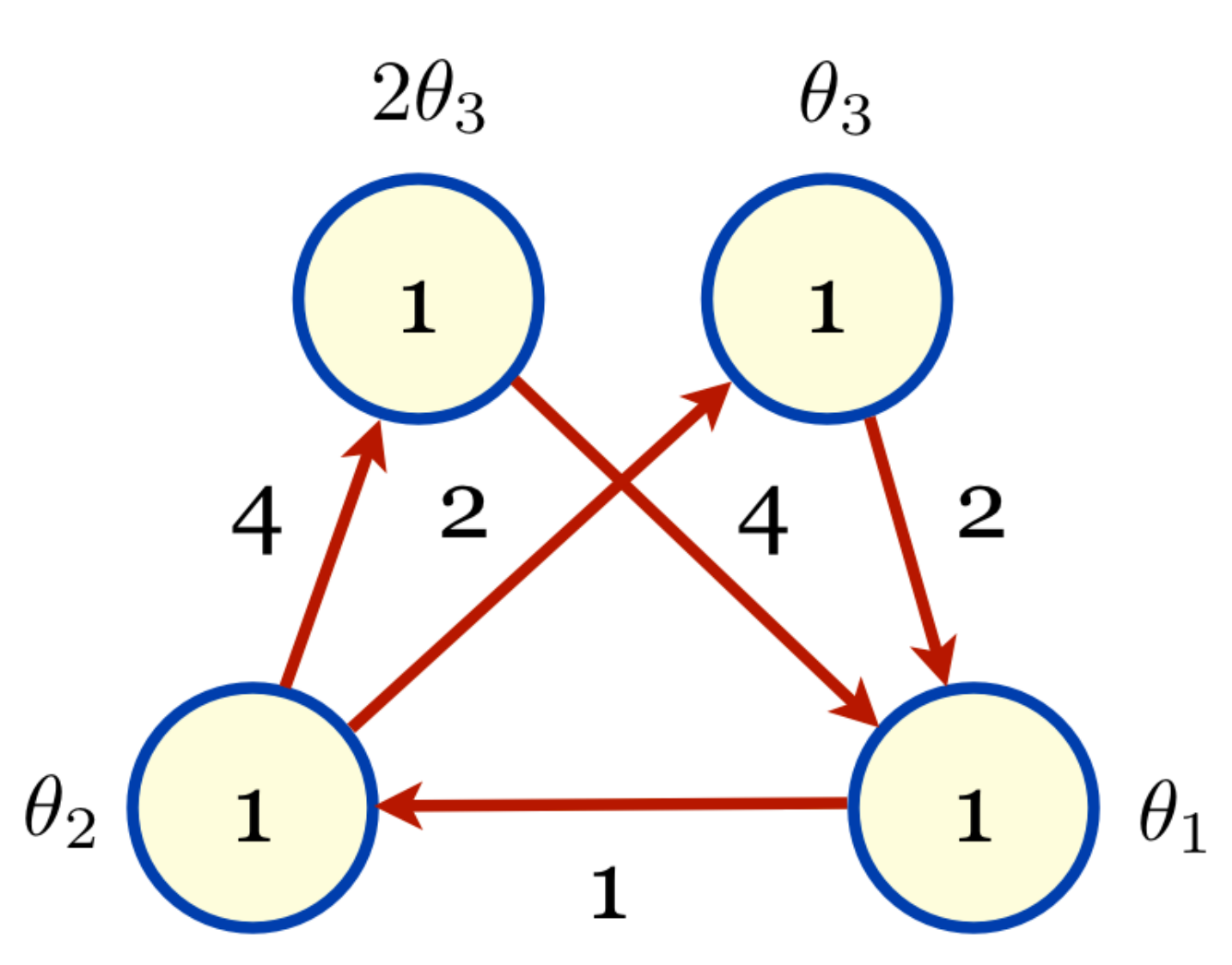}
\includegraphics[scale=0.35]{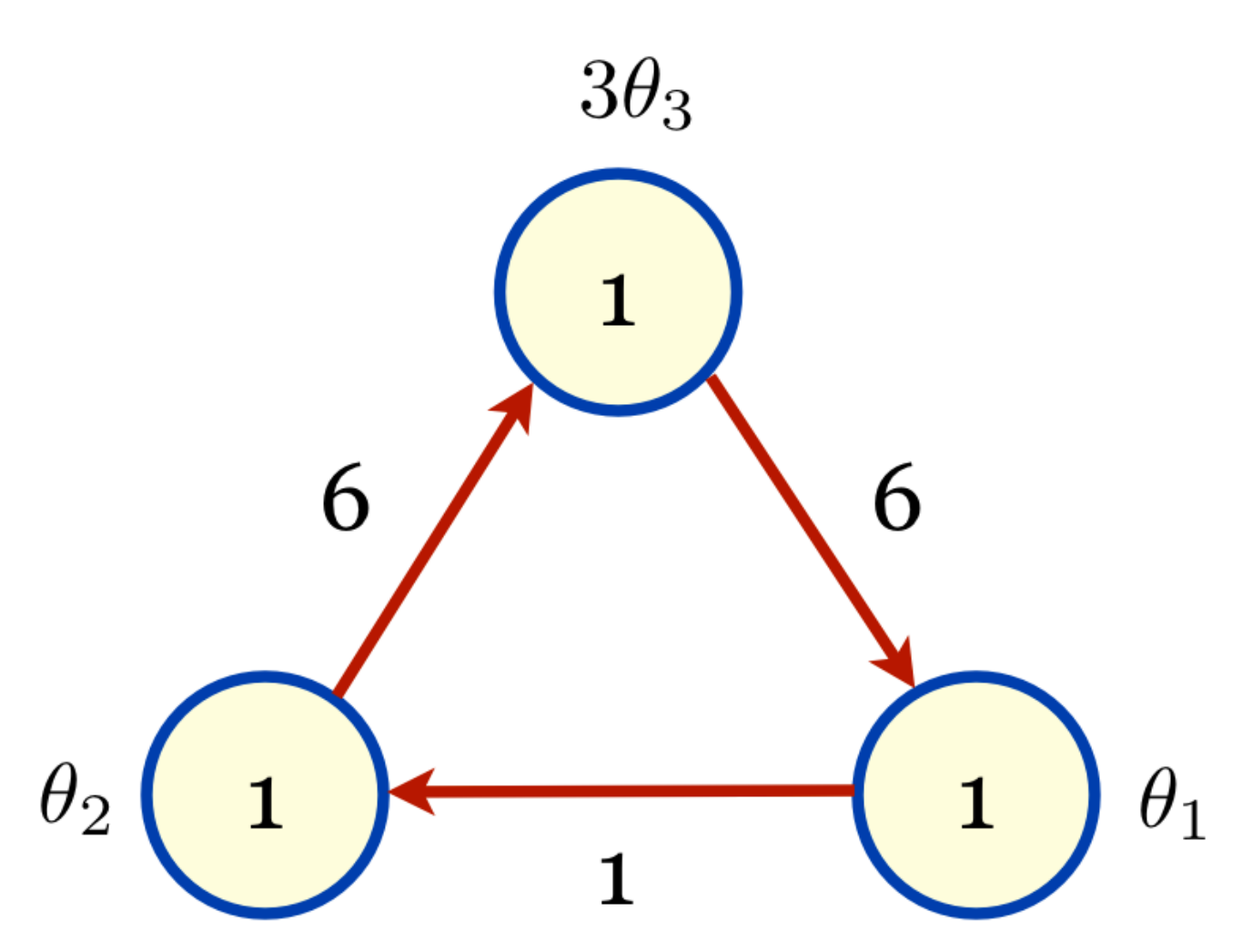}
\caption{The three Abelian quivers arising from the non-Abelian quiver in Fig.~\ref{f:NonAbPartitionEx}; they correspond, respectively(from left to right), to the three partitions $\cP_i$ for $i=1,2,3$ given in Eq.~\eqref{3partitions}.  }\label{f:NonAbPartitionExAbelianized}
\end{figure}

We consider the branch where $\theta_1 > 0$
and $\theta_2 <0$ so that the single bi-fundamental field from node 1 to node 2 vanishes.
Following the Abelianization prescription, we first obtain the
Abelianized quiver; see the first diagram in Fig.~\ref{f:NonAbPartitionExAbelianized}.
Although the corresponding D-term variety $\tilde X$ is still toric, its intersection
structure is not as trivial as that of projective spaces (or products thereof).
Now, the three quivers in Fig.~\ref{f:NonAbPartitionExAbelianized} correspond, respectively, to the three partitions
\bea \label{3partitions}\nn
\cP_1&=&(\{1\}; \{1\}; \{1,1,1\}) \ ,  \\
\cP_2&=&(\{1\}; \{1\}; \{1,2\}) \ , \\ \nn
\cP_3&=&(\{1\};\{1\}; \{3\}) \ , \nn
\eea
 which, in turn, can be obtained from the subsets
\bea \nn
\cI_1 &=& \slashed o \ , \\
\cI_2 &=& \{\alpha_{12}\}, \{\alpha_{13}\}, \{\alpha_{23}\} \ , \\ \nn
\cI_3 &=& \{\alpha_{12}, \alpha_{13}\}, \{\alpha_{12}, \alpha_{23}\}, \{\alpha_{13}, \alpha_{23}\}, \{\alpha_{12}, \alpha_{13}, \alpha_{23}\} \ ,
\eea
of the positive-root set $\Delta^+=\{\alpha_{12}, \alpha_{13}, \alpha_{23}\}$. Note that several subsets $\cI \subset \Delta^+$ can correspond to a given partition $\cP$ in general.

In order to verify the desired partition-sum structure, the two sides of
Eq.~\eqref{pquiver} need to be computed for every partition of $\bold d = (1,1,3)$.
We have indeed computed the left-hand-sides, using our Abelianization procedure, and
find that these match precisely the right-hand-sides, respectively.
For the record, we list them here;
\bea \nn
c({\mathcal P}_1;y)\times \Omega(y)[Q_{{\cal P}_1}] &=& - \frac16 y^{-7} - \frac23 y^{-5} - \frac76 y^{-3} - \frac43 y^{-1} - \frac43 y - \frac 76 y^3 - \frac23 y^5 - \frac 16 y^7 \ ,  \\
c({\mathcal P}_2;y)\times \Omega(y)[Q_{{\cal P}_2}] &=& \frac12 y^{-7} + y^{-5} + \frac32 y^{-3} + 2 y^{-1} + 2 y + \frac32 y^3 + y^5 + \frac12 y^7 \ ,  \\ \nn
c({\mathcal P}_3;y)\times \Omega(y)[Q_{{\cal P}_2}]&=& -\frac13y^{-7} - \frac13y^{-5} -\frac13 y^{-3} - \frac23 y^{-1} - \frac23 y - \frac13 y^3 - \frac13 y^5 - \frac13 y^7  \ .
\eea
The conjectured relation~\eqref{pquiver} is thus confirmed for this cyclic example.

\vskip 1cm

\centerline{\bf Acknowledgement}\noindent
We are grateful to Jaemo Park for collaboration at an early stage
of this work and to Bumsig Kim for kind explanation of his works. We
also thank Tarig Abdelgadir, Jan Manschot, Ashoke Sen, and Dan Xie for useful conversations. P.Y.
is grateful to Institute for Advanced Study, Columbia University, and Perimeter
Institute for hospitality while this work was in progress. This work is supported
in part (P.Y.) by the National Research Foundation of Korea
(NRF) via the Center for Quantum Spacetime (grant number 2005-0049409),
and also (Z.L.W.) by National Natural Science Foundation of China with grant No.11305125.

\appendix

\section{The Ambient D-Term Variety and Maximal Reduced Quivers}\label{Appendix}

For quivers with (oriented) loops, i.e. with superpotentials, the
Higgs moduli space is determined by combining D-term and F-term
constraints. Since the number of F-terms equals to the number of
chiral fields participating in the superpotential, one might think
that the vacua are at most point-like as far as these chiral fields are
concerned, or more likely null if D-terms impose
further nontrivial condition. However, this is not true; there typically
exists Higgs branches with nonnegative dimensions, which arises by setting
entire bi-fundamentals between certain pairs of nodes identically zero.
This kills many F-terms so effectively that the naive dimension counting
above becomes irrelevant.

When this happens (and this happens generically for quivers with
loops), the Higgs moduli space $M$ is embedded, via the surviving
F-term conditions, in certain D-term quiver variety $X$. The quiver
whose Higgs moduli space is $X$ is related to the original quiver
by removal of certain edges so that the final quiver has no loop whatsoever.
$X$ defines the ambient variety we have used throughout this paper.
For the purpose of this Appendix, we call these quivers without loops,
obtained from the original quiver by removal of edges, ``reduced quivers."
We will presently claim that, with generic superpotentials and FI constants,
the reduced quiver has to be always maximal in that restoration of
any one edge would reintroduce a loop. This shows that $X$ is
always a quiver variety obtained by symplectic reduction of higher
dimensional flat complex vector space, $\mathbb{C}^K$, where $K$ is
the total number of chiral superfields in the maximal reduced quiver.

In this Appendix, we shall use $\vec\Phi^{{s}\bar {t}}
=\{\Phi^{a_{s}\bar b_{t}}_{i_{s\bar t}}\}$ denoting the bi-fundamental
fields associated with the edge connecting the node $s$ and $t$, where
$a_{s}$ and $\bar b_{t}$ are the gauge indices and $i_{s\bar t}$ is the flavor index.

Firstly, let us recall the argument in Ref.~\cite{Lee:2012sc} for the Abelian cyclic quivers.
For an $N$-node Abelian cyclic quiver,  we only have the $N$ complex
vectors $\vec\Phi^{s}\equiv \vec\Phi^{s,\overline{s+1}}$ for $s=1, \cdots, N$,
with $\vec \Phi^N = \vec \Phi^{N, \overline{N+1}} \equiv \vec\Phi^{N,\overline 1}$
understood. 
We can show that there is no solution to F-term conditions with
all the complex vectors  nontrivial. If there were such a solution,
it has to be a discrete solution since the number of F-term equations
equals the total number of complex variables $\vec\Phi^{s}$'s if none of
the $\vec\Phi^{s}$ is vanishing. (We always assume that
the coefficients in the superpotential $W=W(\{\Phi^s_{i_s}\})$ are
generic, so the algebraic equations $\partial W=0$
are also generic.) However, the above F-term conditions have $N$
scaling symmetries under $\vec\Phi^{s}\rightarrow \lambda_s\vec\Phi^{s}$
for any complex numbers $\lambda_s$, and one can actually generate $N$
complex dimensional family of solutions. This contradicts the
expected discreteness of the solution, so we cannot generically
expect to find solutions of this type.

Combining with the D-term conditions, one can show \cite{Lee:2012sc}
the ambient D-term variety in each branch is related to a certain
reduced quiver with one edge eliminated.\footnote{By abuse of notation,
in this Appendix, an edge means the whole bunch of arrows between a pair of nodes.}
In general, the eliminated
edge is uniquely decided if the FI parameters are given. The removal
of more than one edges only happens on the marginal stability wall.

For the Abelian multi-loop case, one can show similarly that the ambient D-term variety
is determined entirely by a reduced quiver without loops.
As before, if there is a solution to F-term conditions with all the complex vectors
nonvanishing, it has to be a discrete solution for a generic choice of
the superpotential since the number of F-term equations
equals the total number of complex variables $\vec\Phi^{s\bar t}$'s
which appear in the superpotential. Since the full set of F-term
equations are not homogenous now, we can not use the previous argument to generated a family of solutions.
However, for the quivers we are interested in, there is a $U(1)_R$
symmetry under which the superpotential has charge 2. This
$U(1)_R$ shall generate a one dimensional family of solutions and make
a contradiction to the
expected discreteness. Therefore, at
least one of the edges should be removed. If there are still loops left after
taking one edge to be zero, one can substitute the removed edge into
the superpotential $W$ and repeat the previous argument again and again
until there is no oriented loop left.

\begin {figure}[t]
\centering
\includegraphics[scale=0.38]{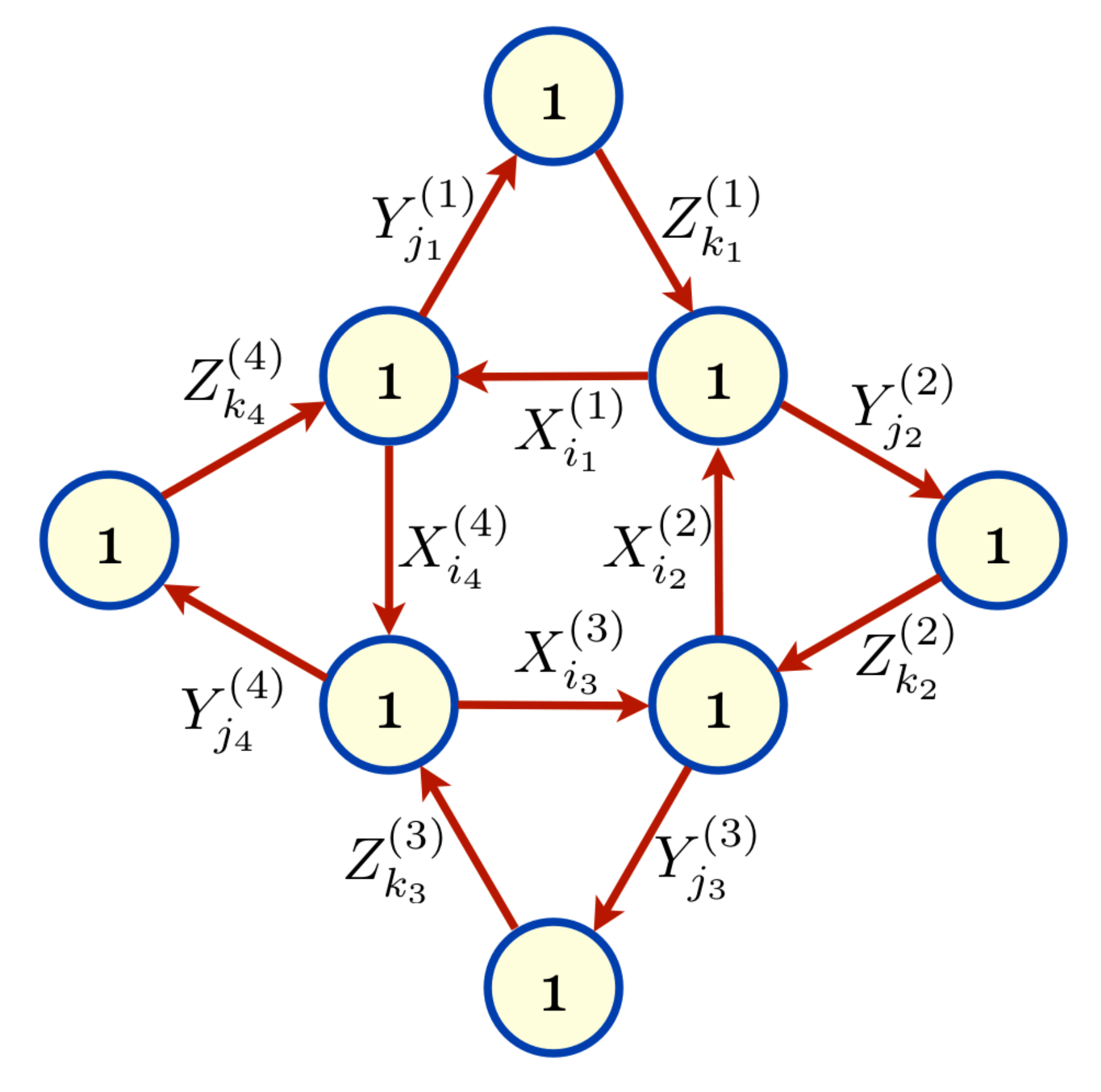}
\caption{Abelian quiver with total of 6 loops; the linking numbers and the
FI constants are implicit, while the bi-fundamental fields are explicitly shown on the edges.}
\label{f:app1}
\end{figure}

For more general cases, the $U(1)_R$
symmetry can be borken. E.g., for the quiver in Fig.~\ref{f:app1}
endowed with a generic superpotential,
\bea \nn
W&=& a_{i_1 i_4 i_3 i_2} X_{i_1}^{(1)} X_{i_4}^{(4)} X_{i_3}^{(3)} X_{i_2}^{(2)} +
\sum\limits_{s=1}^4 b_{i_s j_s k_s}^{(s)} X_{i_s}^{(s)} Y_{j_s}^{(s)} Z_{k_s}^{(s)}  \\ \nn
&&+ c_{j_1 k_1 j_2 k_2 j_3 k_3 j_4 k_4} Y_{j_1}^{(1)} Z_{k_1}^{(1)} Y_{j_2}^{(2)}
Z_{k_2}^{(2)} Y_{j_3}^{(3)} Z_{k_3}^{(3)} Y_{j_4}^{(4)} Z_{k_4}^{(4)}   \ ,
\eea
one can not consistently assign the same $U(1)_R$ for every loop since there are relations among them.
In such cases, we could make the following argument.
Supposing that there is a unbroken loop with $N$ nodes in the quiver, let us
label the corresponding nodes by $1,2,\dots, N$,
and the corresponding edges by $\vec\Phi^{u}$, $u=1,\dots , N$.
We can always construct an action of the form
\bea \nn
\vec\Phi^{u}&\rightarrow& \lambda_u\vec\Phi^{u} \ , \quad \text{for} \quad u=1,\dots ,N-1 \ , \\ \nn
\vec\Phi^{N}&\rightarrow& \left(\prod_{u=1}^{N-1}\lambda_u^{-1}\right)\vec\Phi^{N} \ ,
\eea
to be a symmetry of the F-term equations by also assigning proper scalings
$$\vec\Phi^{s\bar t}\rightarrow \lambda_{s\bar t}(\{\lambda_u\}_{u=1, \cdots, N})\vec\Phi^{s\bar t}$$
  to the edges which are not in this loop.
Again, it means that one can generate $N-1$ complex dimensional family of solutions
and it contradicts the expected discreteness of the solution. The same argument can be repeated
until there is no oriented loop left. Thus, in order to have a non-zero solution in the general Abelian multi-loop case,
there must be an eliminated edge in any single loop.

\begin {figure}[h]
\centering
\includegraphics[scale=0.38]{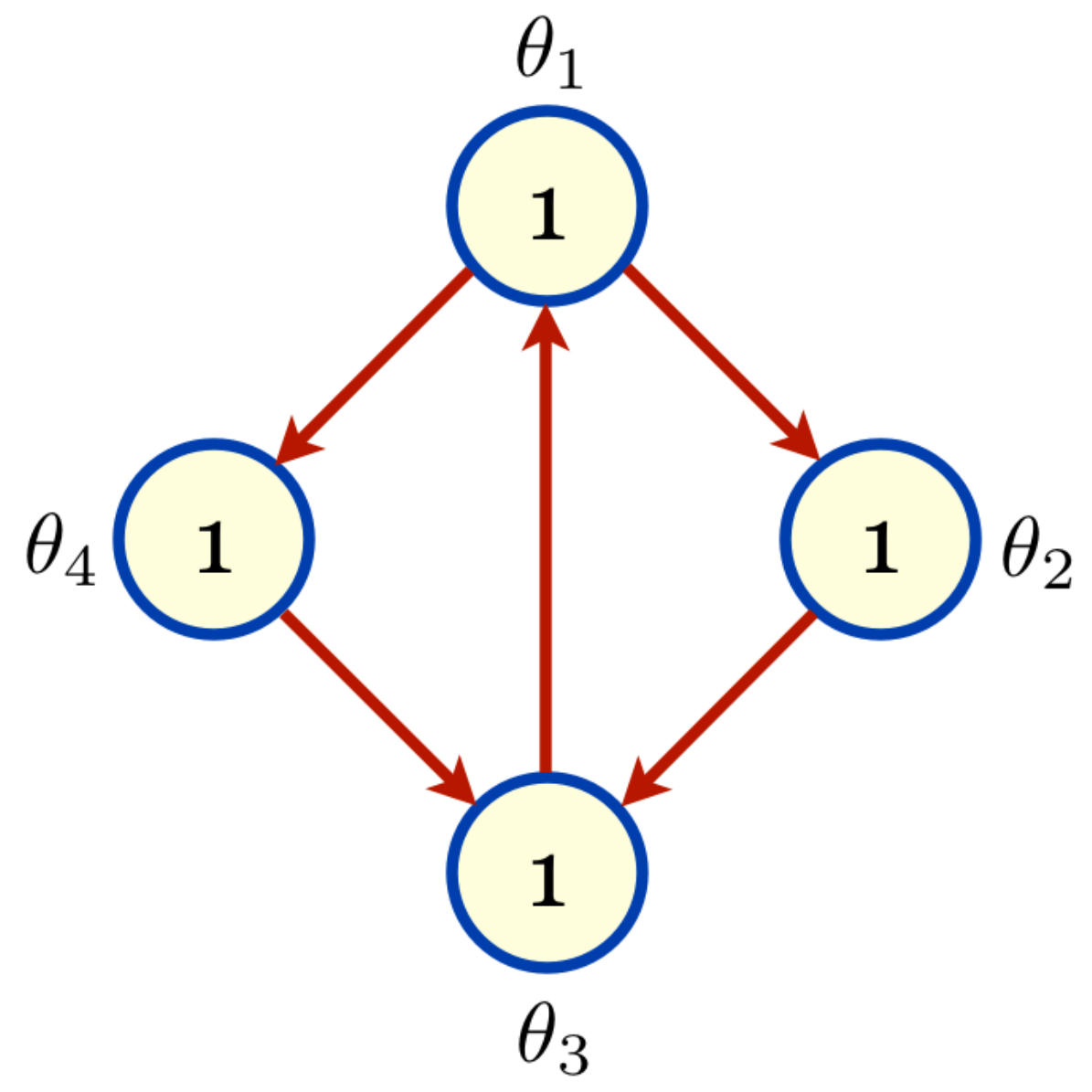}~~~~~
\includegraphics[scale=0.38]{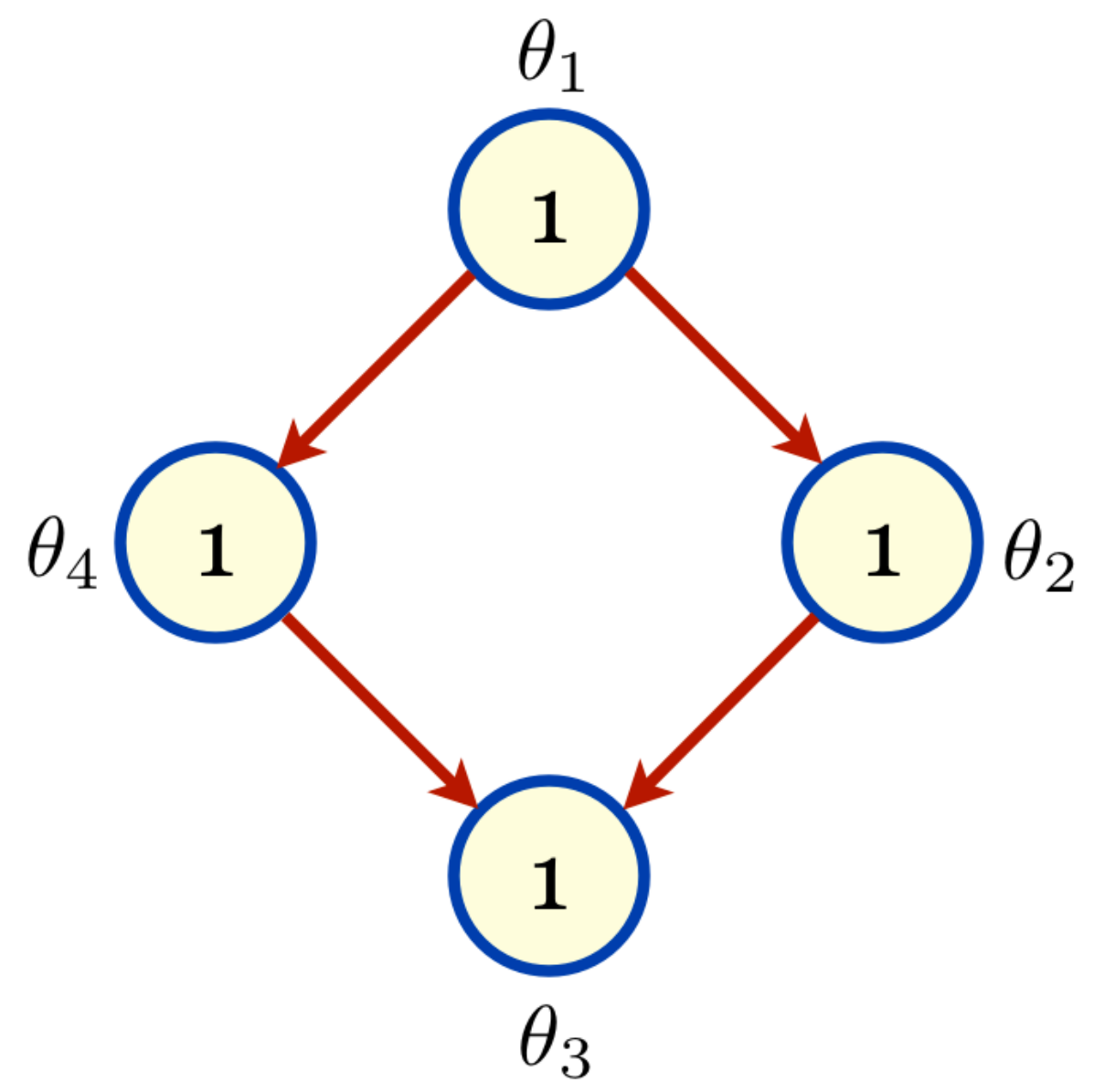}
\caption{Abelian quiver with two loops (left), and one of its maximal
reduced quivers without loop (right) obtained by eliminating the diagonal
edge from node 3 to 1; the linking numbers are implicit. }
\label{f:app2and3}
\end{figure}

Given such removal of edges in a multi-loop quiver, we can derive a
set of consistent conditions on the FI parameters. Unlike the cyclic
case, these constraints could not fix the branch uniquely in general.
That is because the reduced quiver itself could have multiple
non-empty branches depend on more refined choices of FI parameters. E.g.,
for the four-node two-loop quiver in Fig.~\ref{f:app2and3} (left),
by removing the diagonal edge from node 3 to 1, one can obtain the maximal
reduced quiver without loop, depicted in Fig.~\ref{f:app2and3} (right), where,
by ``maximal," we  mean that the reduced quiver will have
a loop if any one of the removed edges, which in this case is the diagonal edge, is recovered.
It is easy to see that removal of this diagonal edge implies the following
consistent conditions
\begin{equation}
\theta_3>0\,,~~~\theta_1<0\,.\nn
\end{equation}
However, this reduced quiver may still have four different, non-trivial,
branches depending on the signs of $\theta_2$ and $\theta_4$.

Thus, the branches for a  Abelian multi-loop quiver are described by
the non-empty branches of its maximal reduced quiver without loop.
Given the maximal reduced quiver without loop
and FI parameters that are read off from the original quiver,
we get the D-term ambient space $X$ in the
corresponding branch. The physical moduli space $M$ is decided
by an intersection of the ambient space variables, and the
intersecting is described by the F-term equations of the removed edges.

The main question for non-Abelian quiver is  whether
the F-term equations are still generic enough for the
components $\Phi^{a_{s}\bar b_{t}}_{i_{s\bar t}}$'s. By ``genericity''
we mean that the equations are algebraically
independent. Although the coefficients of the F-term equation
for $\Phi^{a_{s}\bar b_{t}}_{i_{s\bar t}}$ get repeated for the same
$\{s,t,i_{s \bar t }\}$, they should remain algebraically independent since
the variables are different for different choices of $\{a_{s},\bar b_{t}\}$.
Based on the fact that the F-term equations are still generic
enough for the $\Phi^{a_{s}\bar b_{t}}_{i_{s\bar t}}$'s, the
same argument as the Abelian case will be still valid. Thus,
given an arbitrary quiver, the D-term ambient space $X$ should
be described by the maximal reduced quiver without loop.
This conclusion is consistent with the computation procedure
 we used in the main text.




\begin{thebibliography}{99}
\bibitem{Martin}
S.~Martin,
``Symplectic quotients by a nonAbelian group and by its maximal torus''
[math.SG/0001002].
\end{thebibliography}
\end{document}